\documentclass[journal=jpccck,manuscript=letter,layout=traditional]{achemso}

\usepackage{chemformula} 
\usepackage{textgreek} 
\usepackage{xcolor} 
\usepackage{booktabs} 
\usepackage{multirow} 

\usepackage[separate-uncertainty=true,separate-uncertainty,multi-part-units=single,per-mode=symbol]{siunitx} 
\DeclareSIUnit{\angstrom}{\text{Å}} 
\DeclareSIUnit{\atm}{\text{atm}} 
\DeclareSIUnit{\kcal}{\text{kcal}} 
\DeclareSIUnit{\mol}{\text{mol}} 

\newcommand{\oldspecies}{\ch{I}/\ch{Pb}/\ch{Cs}}
\newcommand{\newspecies}{\ch{I}/\ch{Br}/\ch{Pb}/\ch{Cs}}

\newcommand{\thetax}{\theta_{\mathrm{x}}}
\newcommand{\thetay}{\theta_{\mathrm{y}}}
\newcommand{\thetaz}{\theta_{\mathrm{z}}}

\author{Mike Pols}
    \affiliation{Materials Simulation \& Modelling, Department of Applied Physics and Science Education, Eindhoven University of Technology, 5600 MB, Eindhoven, The Netherlands}
    \alsoaffiliation{Center for Computational Energy Research, Department of Applied Physics and Science Education, Eindhoven University of Technology, 5600 MB, Eindhoven, The Netherlands}
\author{Adri C.T. van Duin}
    \affiliation{Department of Mechanical Engineering, Pennsylvania State University, University Park, PA 16802, United States}
\author{Sof\'{i}a Calero}
    \affiliation{Materials Simulation \& Modelling, Department of Applied Physics and Science Education, Eindhoven University of Technology, 5600 MB, Eindhoven, The Netherlands}
    \email{s.calero@tue.nl}
\author{Shuxia Tao}
    \affiliation{Materials Simulation \& Modelling, Department of Applied Physics and Science Education, Eindhoven University of Technology, 5600 MB, Eindhoven, The Netherlands}
    \alsoaffiliation{Center for Computational Energy Research, Department of Applied Physics and Science Education, Eindhoven University of Technology, 5600 MB, Eindhoven, The Netherlands}
    \email{s.x.tao@tue.nl}

\title{Mixing \ch{I} and \ch{Br} in Inorganic Perovskites: Atomistic Insights from Reactive Molecular Dynamics Simulations}


\begin{document}

\begin{abstract}

All-inorganic halide perovskites have received a lot of attention as attractive alternatives to overcome the stability issues of hybrid halide perovskites that are commonly associated with organic cations. To find a compromise between the optoelectronic properties of \ch{CsPbI3} and \ch{CsPbBr3}, perovskites with \ch{CsPb(Br_{x}I_{1-x})_{3}} mixed compositions are commonly used. An additional benefit is that, without sacrificing the optoelectronic properties for applications such as solar cells or LEDs, small amounts of \ch{Br} in \ch{CsPbI3} can prevent the inorganic perovskite from degrading to a photoinactive nonperovskite yellow phase. Despite indications that strain in the perovskite lattice plays a role in the stabilization of the material, a full understanding of such strain is lacking. Here we develop a reactive force field (ReaxFF) for perovskites starting from our previous work for \ch{CsPbI3}, we extend this force field to \ch{CsPbBr3} and mixed \ch{CsPb(Br_{x}I_{1-x})_{3}} compounds. This force field is used in large-scale molecular dynamics simulations to study perovskite phase transitions and the internal ion dynamics associated with the phase transitions. We find that an increase of the \ch{Br} content lowers the temperature at which the perovskite reaches a cubic structure. Specifically, by substituting \ch{Br} for \ch{I}, the smaller ionic radius of \ch{Br} induces a strain in the lattice that changes the internal dynamics of the octahedra. Importantly, this effect propagates through the perovskite lattice ranging up to distances of \SI{2}{\nm}, explaining why small concentrations of \ch{Br} in \ch{CsPb(Br_{x}I_{1-x})_{3}} (x $\leq$ 1/4) have a significant impact on the phase stability of mixed halide perovskites.

\end{abstract}


\section{Manuscript}

\subsection{Introduction}

Halide perovskites hold a great promise for a variety of optoelectronic applications, which include photovoltaics, light-emitting diodes (LEDs) and photodetectors~\cite{greenEmergencePerovskiteSolar2014, liuMetalHalidePerovskites2021, wangLowDimensionalMetalHalide2021, wuEvolutionFutureMetal2021}. The main appeal of halide perovskites stems from the combination of facile synthesis methods and a highly tunable \ch{AMX3} perovskite crystal lattice~\cite{nayakMechanismRapidGrowth2016, mcmeekinCrystallizationKineticsMorphology2017}. By changing or mixing the \ch{A}-site cation (\ch{MA^{+}}, \ch{FA^{+}}, \ch{Cs^{+}}), \ch{M}-site metal cation (\ch{Pb^{2+}}, \ch{Sn^{2+}}) and \ch{X}-site halide anion (\ch{I^{-}}, \ch{Br^{-}}, \ch{Cl^{-}}), a large compositional space with varying material properties can be explored for specific applications~\cite{jacobssonExplorationCompositionalSpace2016, salibaCesiumcontainingTripleCation2016}. Despite these beneficial material characteristics, the commercialization of perovskite optoelectronic devices has thus far been hampered by long-term stabiliy issues~\cite{zhuMetalHalidePerovskites2018, kunduSituStudiesDegradation2020}. Several of these stability issues, such as a poor thermal stability and material decomposition upon contact with water, can be attributed to the volatile and hydrophilic nature of commonly incorporated organic \ch{A}-site cations (\ch{MA+} and \ch{FA+})~\cite{coningsIntrinsicThermalInstability2015, yangInvestigationCH3NH3PbI3Degradation2015, juarez-perezThermalDegradationFormamidinium2019}.

One strategy that has been proposed to overcome the stability issues related to organic cations, is the use of all-inorganic halide perovskites. Such all-inorganic halide perovskites, in which \ch{Cs+} is the sole \ch{A}-site cation, have shown to be more resistant to external stimuli such as thermal stress and moisture~\cite{xiangReviewStabilityInorganic2021}. As a result of this, they have been used in a variety of applications. For example, \ch{CsPbI3}, with its relatively low band gap (\SI{1.73}{\eV}~\cite{eperonInorganicCaesiumLead2015}) is ideal for solar cells~\cite{duanPhasePureGCsPbI3Efficient2022} and LEDs emitting red light~\cite{chenAminoAcidPassivatedPure2023}. Whereas \ch{CsPbBr3}, with its larger band gap (\SI{2.37}{\eV}~\cite{manninoTemperatureDependentOpticalBand2020}), is commonly used in tandem solar cells~\cite{liAllInorganicCsPbBr3Perovskite2019}, green LEDs~\cite{wangTrifluoroacetateInducedSmallgrained2019} and photodetectors~\cite{stoumposCrystalGrowthPerovskite2013}. Moreover, such all-inorganic perovskites can be tuned through nanostructuring, offering improved stability and optoelectronic properties for an even wider range of applications~\cite{protesescuNanocrystalsCesiumLead2015}.

Nevertheless, all-inorganic perovskites are not without any problems, as indicated by the poor phase stability of \ch{CsPbI3}. It is well established that \ch{CsPbI3} transforms from a cubic (\textalpha) to tetragonal (\textbeta) to orthorhombic (\textgamma) phase, going from high to progressively lower temperatures~\cite{stoumposRenaissanceHalidePerovskites2015, marronnierAnharmonicityDisorderBlack2018}. Due to a mismatch of the ionic radii in the lattice, evidenced by the low Goldschmidt tolerance factor of \ch{CsPbI3} (0.807)~\cite{goldschmidtGesetzeKrystallochemie1926, shannonRevisedEffectiveIonic1976}, the low-temperature \textgamma-phase is rather distorted, making it prone to convert into a nonperovskite yellow (\textdelta) phase~\cite{stoumposRenaissanceHalidePerovskites2015, marronnierAnharmonicityDisorderBlack2018, strausUnderstandingInstabilityHalide2020}. The yellow phase of \ch{CsPbI3} is photoinactive, which is ill-suited for optoelectronic applications. On the contrary, resulting from the better fit of the ions in the lattice indicated by its higher Goldschmidt tolerance factor (0.815)~\cite{goldschmidtGesetzeKrystallochemie1926, shannonRevisedEffectiveIonic1976}, \ch{CsPbBr3} does not show any degradation to a yellow phase. Therefore, \ch{I-} and \ch{Br-} ions are commonly mixed to increase the phase stability of all-inorganic perovskites. A variety of works have demonstrated, that apart from impacting the optoelectronic properties~\cite{bealCesiumLeadHalide2016, suttonBandgapTunableCesiumLead2016, ghoshDependenceHalideComposition2018}, the mixing of \ch{Br} into a \ch{CsPbI3} enhances the stability of \ch{CsPb(Br_{x}I_{1-x})_{3}} films, either slowing down or preventing the yellow phase from forming altogether~\cite{suttonBandgapTunableCesiumLead2016, chenAllVacuumDepositedStoichiometricallyBalanced2017, ghoshDependenceHalideComposition2018, liCsBrInducedStableCsPbI32018, linThermochromicHalidePerovskite2018, liuCsPbI3PerovskiteQuantum2021}. \citet{nasstromDependencePhaseTransitions2020} systematically studied the phase transitions of \ch{CsPb(Br_{x}I_{1-x})_{3}} perovskites, from which they found that a gradual increase of the \ch{Br} content in mixed halide perovskites, lowers the temperatures at which the perovskite transforms into the cubic phase. Although lattice strain has been proposed as the mechanism responsible for \textalpha-phase stabilization~\cite{eperonInorganicCaesiumLead2015, liCsBrInducedStableCsPbI32018}, the atomistic effects of halide mixing on the various perovskite phases remains unclear.

Recently, using reactive force field (ReaxFF) molecular dynamics simulations, we studied the phase transitions and degradation reactions at surfaces and grain boundaries of \ch{CsPbI3}~\cite{polsAtomisticInsightsDegradation2021, polsWhatHappensSurfaces2022}. In this work, we extend our study to the lattice and ion dynamics of mixed halide perovskites. Starting from our ReaxFF parameter set for \ch{CsPbI3}~\cite{polsAtomisticInsightsDegradation2021} we expand the force field to \ch{CsPbBr3} and mixed \ch{CsPb(Br_{x}I_{1-x})_{3}} perovskites. After validating the new ReaxFF parameters on the equations of state, mixing enthalpies, degradation reactions and defect migration barriers, we apply our model in large-scale molecular dynamics simulations of mixed perovskites. By combining information from the phase diagrams with the microscopic order in the octahedral orientations, we provide important atomistic insights into the effects of halide mixing.

\subsection{Methods}

We train the ReaxFF parameters for \ch{CsPb(Br_{x}I_{1-x})3} halide perovskites against reference data from density functional theory (DFT) calculations performed in VASP~\cite{kresseInitioMoleculardynamicsSimulation1994, kresseEfficiencyAbinitioTotal1996, kresseEfficientIterativeSchemes1996} and ADF~\cite{fonsecaguerraOrderNDFTMethod1998, teveldeChemistryADF2001} using the PBE+D3(BJ)~\cite{perdewGeneralizedGradientApproximation1996, grimmeEffectDampingFunction2011} exchange-correlation functional. The reference data includes atomic charges, equations of state of different perovskite and nonperovskite phases, equations of state of precursors (e.g. \ch{CsX} and \ch{PbX2} with \ch{X} = \ch{I}/\ch{Br}), defect formation energies, defect migration barriers, and phase transitions of compounds. Full details of these calculations are found in SI Note 1 in the Supporting Information. The agreement between the reference data and the predictions from the ReaxFF parameter set $\{ p_{j} \}$ are captured by a sum of squared errors (SSE) loss function as
\begin{equation}
    \mathrm{SSE} \left( \{ p_{j} \} \right) = \sum^{N}_{i = 1} \left[ \frac{ x_{i\mathrm{,ref}} - x_{i\mathrm{,calc}} \left( \{ p_{j} \} \right) }{\sigma_{i}} \right]^{2}
\end{equation}
where $x_{i\mathrm{,ref}}$ and $x_{i\mathrm{,calc}}$ are the reference values and ReaxFF predictions for an entry $i$ in the training set and $\sigma_{i}$ the weight of that entry. The final ReaxFF parameter set is obtained by minimizing the SSE loss function using the covariance matrix adaptation evolution strategy (CMA-ES)~\cite{hansenCompletelyDerandomizedSelfAdaptation2001} as implemented in ParAMS in AMS2022~\cite{komissarovParAMSParameterOptimization2021, AMS2022}. We use the previously published \oldspecies{} parameters~\cite{polsAtomisticInsightsDegradation2021} for inorganic halide perovskites as the initial point for parameter optimization. Without any ReaxFF parameters for \ch{Br} available in literature, the starting point for the \ch{Br} parameters was obtained by scaling the interactions of \ch{I} with other species. Details of the parameter optimization procedure and the scaling of the interatomic interactions can be found in SI Note 2 in the Supporting Information.

\subsection{Results and discussion}

\subsubsection{Force field validation}

\begin{figure*}[htbp!]
    \includegraphics{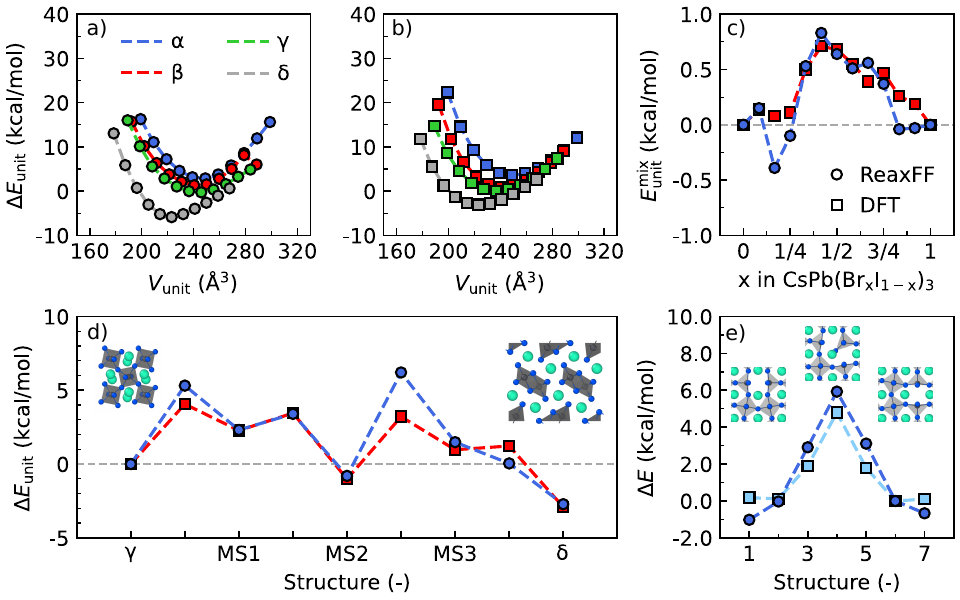}
    \caption{Equations of state of various perovskite and non-perovskite phases of \ch{CsPbI3} from a) ReaxFF and b) DFT calculations. c) Mixing enthalpies of \ch{CsPb(Br_{x}I_{1-x})_{3}} perovskites. d) \ch{CsPbI3} degradation mechanism from the \textgamma-phase to the \textdelta-phase. e) Defect migration barrier of \ch{I} vacancy in \ch{CsPbI3}. Data from ReaxFF simulations and DFT calculations are shown in circles and squares, respectively. Degradation mechanism data from Ref.~\citenum{chenKineticPathwayGtod2022}.}
    \label{fig:parameter_validation}
\end{figure*}

Using the above-mentioned optimization procedure, we obtain a ReaxFF description for the elements \newspecies{} that exhibits a good agreement with the DFT reference data in the training set, as shown in Figure~\ref{fig:parameter_validation} and SI Note 3 in the Supporting Information. The obtained ReaxFF parameter set is provided in the Supporting Information. Focusing on pure compounds first, we note that the equations of state of the various phases of \ch{CsPbI3}, both perovskite (\textalpha-, \textbeta-~and \textgamma-phase) and nonperovskite (\textdelta-phase), as obtained with ReaxFF (Figure~\ref{fig:parameter_validation}a) are in good agreement with DFT calculations (Figure~\ref{fig:parameter_validation}b). We find that the ReaxFF parameter set correctly ranks the total energies of the various bulk phases of \ch{CsPbI3} from least to most stable as: \textalpha~$<$ \textbeta~$<$ \textgamma~$<$ \textdelta. Moreover, in agreement with the reference data, the ReaxFF parameter set predicts a similar stability trend for the different phases of \ch{CsPbBr3}, an overview of which is shown in Figure S1 and Table S5. As shown in Figure~\ref{fig:parameter_validation}c, the ReaxFF force field also predicts positive mixing enthalpies for mixed halide compositions ($<$ \SI{1.0}{\kcal\per\mol} per formula unit) in agreement with DFT calculations. We hypothesize the discrepancies in the mixing enthalpies at x = 1/6 and x = 1/4 can be linked to overstabilized mixed perovskite structures, more details of which are provided in SI Note 3 in the Supporting Information.

Shifting our focus from pristine bulk systems to degradation reactions and defective perovskites, we find that such systems are also represented well by the new ReaxFF parameter set (Figure~\ref{fig:parameter_validation}d). In particular, the new ReaxFF force field captures the energetics of the the degradation of \ch{CsPbI3} from the orthorhombic (\textgamma) to the yellow (\textdelta) phase as predicted by DFT calculations for the structures from \citet{chenKineticPathwayGtod2022}. Compared to the previously published \oldspecies{} parameter set (Figure S2)~\cite{polsAtomisticInsightsDegradation2021}, the reparameterized force field provides considerable improvements for the stability of the metastable states (MS1, MS2 and MS3) and final state (\textdelta) in the degradation pathway, potentially paving the way for the simulation of this degradation reaction using rare event sampling methods. Finally, we find that the ReaxFF force field finds defect migration barriers of halide point defects (i.e. vacancies and interstitials) that are in line with migration barriers from DFT calculations. Figure~\ref{fig:parameter_validation}e demonstrates that the migration of an \ch{I} vacancy in \ch{CsPbI3} from ReaxFF \SI{+4.8}{\kcal\per\mol} is close to that from DFT calculations \SI{7.0}{\kcal\per\mol}. Defect migration barriers of other types of defects, such as an \ch{I} interstitial in \ch{CsPbI3} or a \ch{Br} vacancy or interstitial in \ch{CsPbBr3}, are also correctly predicted by the new ReaxFF force field, an overview of these barriers is shown in Figure S3.

\begin{figure}[htbp!]
    \includegraphics{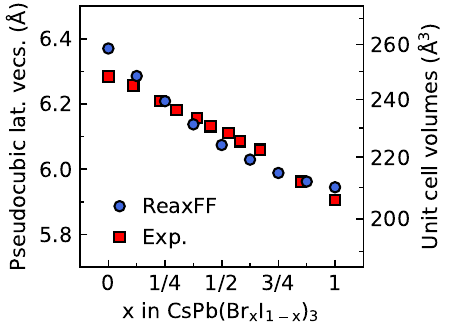}
    \caption{Pseudocubic lattice vectors and unit cell volumes of \ch{CsPb(Br_{x}I_{1-x})_{3}} perovskites. Comparison of experimental data (squares) with ReaxFF simulations (circles) at \SI{575}{\K}. Experimental data from Ref.~\citenum{nasstromDependencePhaseTransitions2020}.}
    \label{fig:unit_cell_volumes}
\end{figure}

To assess the performance of the new parameter set during finite temperature simulations, we compare unit cell volumes from simulations with experimentally observed volumes~\cite{nasstromDependencePhaseTransitions2020} in Figure~\ref{fig:unit_cell_volumes}. The full details of the creation of the model systems and the simulations can be found in SI Note 4 and SI Note 5 in the Supporting Information. Notably, we find that the ReaxFF simulations predict volumes within 1\% of experiments, demonstrating that an increase in the \ch{Br} content in the \ch{CsPb(Br_{x}I_{1-x})_{3}} lattice, decreases the unit cell volumes. An effect that can be attributed to the smaller ionic radius of \ch{Br} (\SI{1.96}{\angstrom}) compared to \ch{I} (\SI{2.20}{\angstrom})~\cite{shannonRevisedEffectiveIonic1976}, which reduces the size of the crystal lattice. 

\subsubsection{Phase diagrams}

\begin{figure*}[htbp!]
    \includegraphics{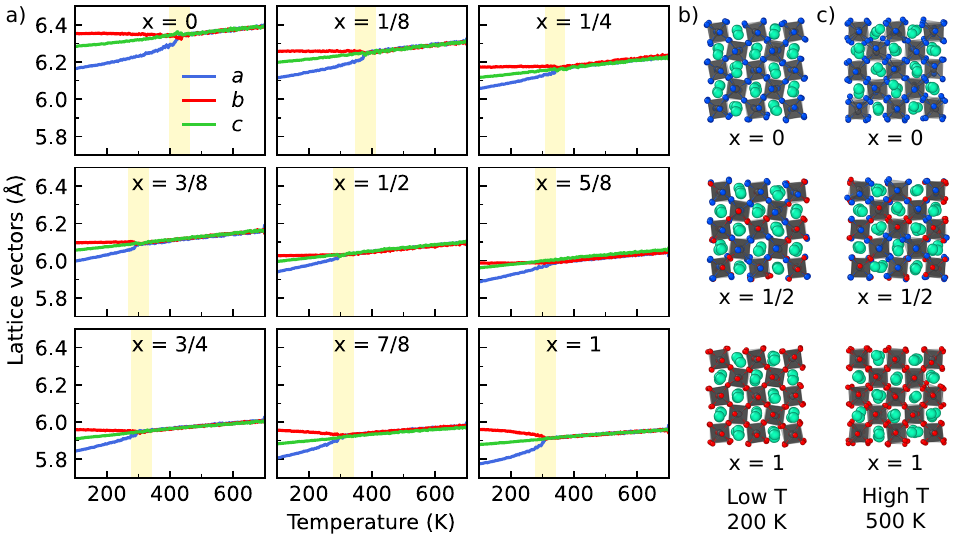}
    \caption{a) Phase diagrams of \ch{CsPb(Br_{x}I_{1-x})_{3}} perovskites with varying compositions obtained during the gradual heating of the inorganic compounds. Snapshots of mixed halide perovskites with x = 0, x = 1/2 and x = 1 compositions are shown in b) \SI{200}{\K} and c) \SI{500}{\K}. The yellow bars indicate the temperature at which the cubic phase is initially observed. The pseudocubic lattice vectors $a$, $b$ and $c$ are shown in all figures.}
    \label{fig:phase_transitions}
\end{figure*}

Having established that the ReaxFF parameter set can appropriately describe the macroscopic properties of mixed compositions at various temperatures, we now shift our focus to studying the transitions between the various perovskite phases. To do so, we gradually heat different \ch{CsPb(Br_{x}I_{1-x})_{3}} systems with varying amounts of \ch{Br} (x = 0, 1/8, 1/4, 3/8, 1/2, 5/8, 3/4, 7/8 and 1), from \SI{100}{\K} to \SI{700}{\K} and monitor the temperature evolution of the lattice vectors in Figure~\ref{fig:phase_transitions}. Details of the simulations used to obtain the phase diagrams can be found in SI Note 5 in the Supporting Information.

Looking into the phase diagrams of the pure perovskites (x = 0 and x = 1 in Figure~\ref{fig:phase_transitions}a), we observe that both perovskites transition from the low-temperature orthorhombic phase to a high-temperature cubic phase. As shown in the snapshots in Figure~\ref{fig:phase_transitions}b and Figure~\ref{fig:phase_transitions}c, the perovskites change from a phase in which the octahedra are arranged in an ordered tilted fashion at low temperatures (\SI{200}{\K}), to one where this order in the octahedral tilting is overcome by the dynamic alternation between many different tilts at high temperatures (\SI{500}{\K}). We distinguish these phaes based on the magnitude of the lattice vectors; in the orthorhombic phase all lattice vectors are different ($a \neq b \neq c$) and in the cubic phase all vectors are the same length ($a = b = c$). The intermediate tetragonal phase ($a = b \neq c$) only appears during a narrow temperature window for pure \ch{CsPbI3} in Figure~\ref{fig:phase_transitions}  (\SI{410}{\K} to \SI{430}{\K}), as a result of rapid thermal fluctuations. In agreement with experiments~\cite{marronnierAnharmonicityDisorderBlack2018, stoumposCrystalGrowthPerovskite2013, nasstromDependencePhaseTransitions2020}, we find that \ch{CsPbBr3} (\SI{310}{\K}) transforms to the cubic phase at lower temperatures compared to \ch{CsPbI3} (\SI{430}{\K}). This difference in the phase transition temperatures indicates that a smaller amount of thermal fluctuations is needed for phase transitions to occur in \ch{CsPbBr3}~\cite{yangSpontaneousOctahedralTilting2017, jinnouchiPhaseTransitionsHybrid2019}, an observation that can be linked to the aforementioned higher Goldschmidt tolerance factor of \ch{CsPbBr3} (0.815) compared to \ch{CsPbI3} (0.807)~\cite{goldschmidtGesetzeKrystallochemie1926, shannonRevisedEffectiveIonic1976}. It should be noted that the phase transition temperatures from ReaxFF underestimate the experimental phase transition temperatures by approximately \SI{50}{\K} to \SI{100}{\K} for both \ch{CsPbBr3} and \ch{CsPbI3}. We relate this overprediciton to the exchange-correlation functional used for the training set (i.e. PBE+D3(BJ)), the choice of which has an impact on the phase transition temperatures~\cite{franssonPhaseTransitionsInorganic2023}.

Focusing on the mixed compositions, we find that the mixing of \ch{Br} into \ch{CsPbI3} significantly lowers the phase transition temperature to the cubic phase of perovskites. Furthermore, the phase diagrams in Figure~\ref{fig:phase_transitions}a show that the largest part of the drop in the phase transition temperature occurs at relatively low \ch{Br} concentrations (x $\leq$ 1/4), levelling off for concentrations from x = 1/4 onwards. This finding is consistent with earlier experimental investigations in which it was also found that the phase transition temperature of mixed halide perovskites depends nonlinearly on the \ch{Br} concentration, with the largest drop occurring for small amounts of \ch{Br}~\cite{nasstromDependencePhaseTransitions2020, sharmaPhaseDiagramsQuasibinary1992}.

\subsubsection{Octahedral dynamics}

\begin{figure*}[htbp!]
    \includegraphics{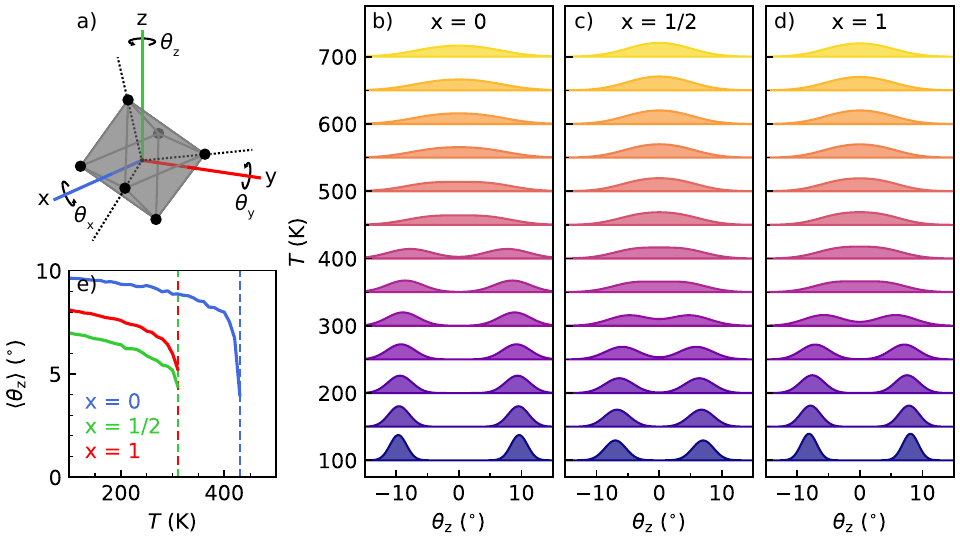}
    \caption{a) Angles $\thetax$, $\thetay$ and $\thetaz$ used to determine the orientation of the \ch{PbX6} octahedra. Temperature evolution of the octahedral orientation $\thetaz$ for \ch{CsPb(Br_{x}I_{1-x})_{3}} perovskites with compositions b) x = 0, c) x = 1/2 and d) x = 1. e) Temperature evolution of the average tilting angle $\langle {\thetaz} \rangle$ for different mixed halide perovskite compositions.}
    \label{fig:octahedral_tilting_distributions}
\end{figure*}

To gain more insights into the phase behavior of \ch{CsPb(Br_{x}I_{1-x})_{3}}, we analyze the orientation of the \ch{PbX6} octahedra in the lattice. Using the method outlined by \citet{wiktorQuantifyingDynamicTilting2023}, the orientation of the octahedra with respect to the cubic lattice can be described by the angles $\thetax$, $\thetay$ and $\thetaz$, following the convention shown in Figure~\ref{fig:octahedral_tilting_distributions}a. The angles act as a measure for the degree with which the octahedra are distorted in the perovskite lattice. We obtain insights into the effects of halide mixing onto the internal dynamics by examining the temperature progression of the octahedral tilting from continuously heated runs at atmospheric pressure for various compositions. The full simulation details and the procedure used to extract the octahedral orientation from the simulations are found in SI Note 5 and SI Note 6 in the Supporting Information.

The temperature evolution of $\thetaz$ is shown in Figure~\ref{fig:octahedral_tilting_distributions}b-d for various compositions, whereas the evolution of $\thetax$ and $\thetay$ can be found in the Figure S10 and Figure S11. All angles, $\thetax$, $\thetay$ and $\thetaz$, change from a bimodal distribution around zero at low temperatures to a single broad distribution centered at zero at high temperatures. This indicates that all compositions progress from a low temperature phase in which the octahedra have a regularly distorted arrangement, to a high-temperature phase that lacks any instantaneous order, but on average has a non-distorted tilting pattern. We note that these observations are in line with with the phase transitions of perovskites where the material progresses from an orthorhombic phase into a cubic phase upon gradual heating, as shown in Figure~\ref{fig:phase_transitions}. Further analysis of the tilting distributions, by means of a symmetric double Gaussian fit (Figure~\ref{fig:octahedral_tilting_distributions}e), allows for the comparison of the various compositions at temperatures ranging from \SI{100}{\K} to \SI{430}{\K}. The comparison illustrates that \ch{PbX6} octahedra in \ch{CsPbI3} have an on average larger tilt than those in \ch{CsPbBr3}. Interestingly, we find that the mixed halide perovskite (x = 1/2) exhibits a smaller average tilt angle and wider tilt distributions for all angles than either of the pure compounds (x = 0 or x = 1), which can be linked to substantial atomistic changes in the perovskite lattice as a result of the halide mixing.

\subsubsection{Atomistic effects of halide mixing}

\begin{figure*}[htbp!]
    \includegraphics{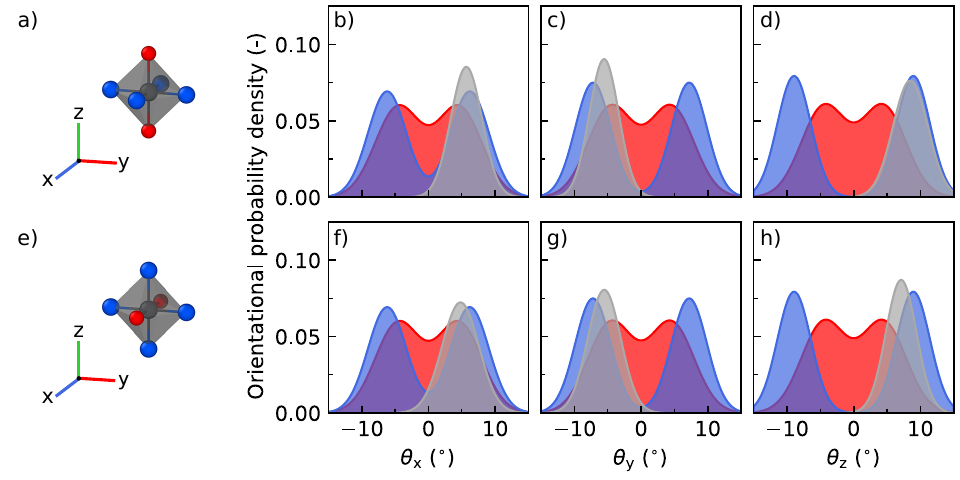}
    \caption{Tilting distributions of \ch{PbX6} octahedra. a) Two \ch{Br} substitutions at the axial position with b-d) showing the distributions of $\thetax$, $\thetay$ and $\thetaz$. e) Two \ch{Br} substitutions at the equatorial position with f-h) showing the distributions of $\thetax$, $\thetay$ and $\thetaz$. The tilting distributions of the substituted octahedra are shown in gray, those for the pure compounds \ch{CsPbI3} and \ch{CsPbBr3} are shown in blue and red, respectively.}
    \label{fig:microscopic_dynamics}
\end{figure*}

Finally, to investigate the atomistic effects of halide mixing, we analyze the tilting distributions of the \ch{PbX6} octahedra in the dilute limit. By mixing small amounts of \ch{Br} into pure \ch{CsPbI3}, we can identify the atomistic effects such substitutions have on the octahedral tilting. In Figure~\ref{fig:microscopic_dynamics} we focus on the tilting distributions of \ch{Br}-substituted \ch{PbI6} octahedra and compare them with tilting distributions of octahedra in pure \ch{CsPbI3} and \ch{CsPbBr3}. To prevent thermal motion from dominating the motion of the octahedra in \ch{CsPbI3}, we investigate the mentioned effects in the low temperature \textgamma-phase at \SI{300}{\K}. In this phase, two types of halide substitutions are possible: 1) axial halide substitutions along the $z$-direction of the octahedra (Figure~\ref{fig:microscopic_dynamics}a) and 2) equatorial substitutions in the $xy$-plane of the octahedra (Figure~\ref{fig:microscopic_dynamics}e). Both types of substitutions are investigated. An overview of the simulation details and model systems used during the simulations can be found in SI Note 5, whereas additional analyses of the octahedral tilting are found in SI Note 7 in the Supporting Information.

The tilting distributions in Figure~\ref{fig:microscopic_dynamics} show that halide substitutions impact the the orientation of octahedra. For both types of substitutions, axial (Figure~\ref{fig:microscopic_dynamics}b-d) and equatorial (Figure~\ref{fig:microscopic_dynamics}f-h), the orientation of the \ch{Br}-substituted \ch{PbI6} octahedra shifts away from that of pure \ch{CsPbI3} to that of \ch{CsPbBr3} by decreasing by about \SI{1}{\degree} to \SI{2}{\degree}. Besides, the octahedral tilting distributions become more narrow. The exact values of the shift and narrowing of the tilting distributions can be found in Table S9. Together, the decreasing tilt angles and the narrowing of the tilting distributions indicate a restrained motion for the substituted octahedra. Specifically, whenever \ch{Br} connects two octahedra in a perovskite lattice that predominantly consists of \ch{PbI6} octahedra, a strained interconnect is formed between the substituted octahedra as a result of the earlier-mentioned smaller size of \ch{Br} compared to \ch{I}, which leads to shorter bond lengths. To alleviate this strain, the substituted octahedra adjust themselves to an overall less tilted geometry, that closely resembles the cubic phase, at low temperatures. This effect is largest for octahedral orientations perpendicular to the substitution direction, for example $\thetax$ and $\thetay$ for axial substitutions. Although the effect of substituting two \ch{Br} into one \ch{PbX6} octahedron is demonstrated here, we note that the substitution of a single \ch{Br} into an octahedron has similar effects as shown in Figure S14.

\begin{figure*}[htbp!]
    \includegraphics{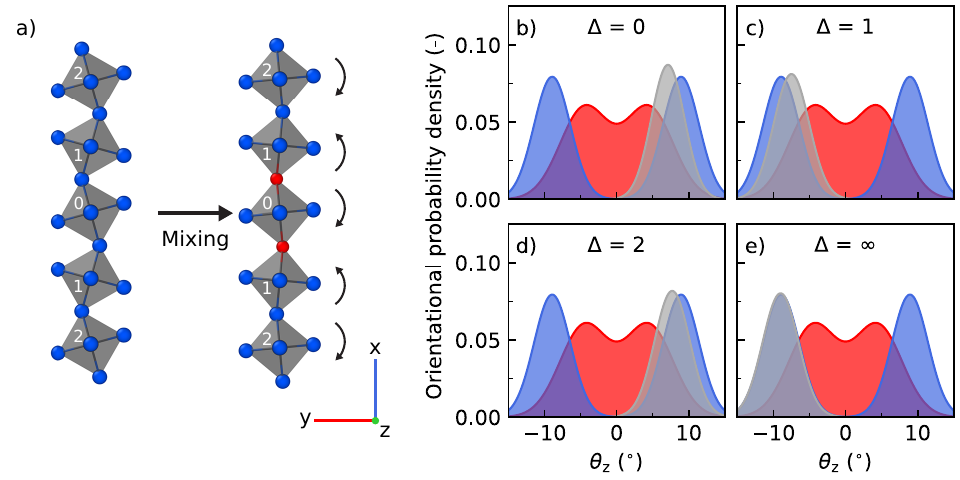}
    \caption{a) Non-substituted and double \ch{Br}-substituted chains of \ch{PbX6} octahedra. The numbers in the octahedra indicate the distance relative to the substituted octahedron. Distribution of $\thetaz$ of b) substituted octahedron ($\Delta = 0$), c) direct neighbor of the substituted octahedron ($\Delta = 1$), d) octahedron two sites away from the substituted octahedron ($\Delta = 2$) and e) reference octahedron very far away from the halide substitution ($\Delta = \infty$). The tilting distributions of the investigated octahedra are shown in gray, those for \ch{CsPbI3} and \ch{CsPbBr3} in blue and red, respectively.}
    \label{fig:substitution_propagation}
\end{figure*}

To investigate the range of the effect of halide substitutions, we monitor the octahedral tilting of octahedra close to an octahedron with two equatorial substitutions in Figure~\ref{fig:substitution_propagation}. A schematic overview of the octahedra that were considered is shown in Figure~\ref{fig:substitution_propagation}a. We find that the tilting distributions of the octahedra close to the substitution (Figure~\ref{fig:substitution_propagation}b-d) deviate from the tilting distributions observed in pure \ch{CsPbI3}. The affected octahedra show a smaller average angle and more narrow distribution for $\thetaz$ as shown in Table S10. The effect diminishes for octahedra far away from the halide substitutions (Figure~\ref{fig:substitution_propagation}e), becoming negligible for octahedra spaced further than three sites away from the substitution ($\Delta > 3$) as shown in Figure S15 and Table S11. We identify the propagation distance of this effect to be about \SI{2}{\nm}. Interestingly, this propagation is not only found in the direction of the halide substitution as shown in Figure~\ref{fig:substitution_propagation}, but also in directions perpendicular to the substitutions, albeit at a shorter range ($<$ \SI{1}{\nm}) as seen in Figure S16 and Table S12. As a consequence of the propagation of this effect, small concentrations of halide substitutions can have profound effects on the octahedral dynamics of perovskites. These atomistic insights are important for understanding why low levels of \ch{Br} (x $\leq$ 1/4) are sufficient to stabilize the cubic phase in \ch{CsPb(Br_{x}I_{1-x})_{3}} perovskites.

\subsection{Conclusion}

In summary, we developed a \newspecies{} ReaxFF parameter set for inorganic halide perovskites. We demonstrate that the developed force field is suitable for describing the various perovskite and nonperovskite phases of pure \ch{CsPbI3}, pure \ch{CsPbBr3} and mixed \ch{CsPb(Br_{x}I_{1-x})_{3}} compounds. By studying the phase transitions of \ch{CsPb(Br_{x}I_{1-x})_{3}} perovskites, we find that progressively increasing the \ch{Br} content stabilizes the high-temperature cubic phase. We highlight that a large part of the stabilization effect comes from the initial \ch{Br} substitutions (x $\leq$ 1/4). An investigation of the octahedral tilting distributions in mixed perovskites shows that halide mixing induces strain in the lattice, causing the perovskite to adopt a more cubic structure. Importantly, the effect of this strain propagates to octahedra close to the substitution, reaching distances of up to \SI{2}{\nm}. These results provide fundamental insights into the microscopic effects of strain that result from halide mixing, which are valuable in the development of optoelectronic devices based on inorganic halide perovskites. Finally, we expect the newly developed ReaxFF parameters to also be used to study other important phenomena, such as defect migration and degradation reactions, occurring in inorganic mixed halide perovskites of various dimensions (e.g. 2D and quantum dots) with large-scale molecular dynamics simulations.


\begin{suppinfo}

\begin{itemize}
    \item Computational settings of the training set entries; parameter optimization procedure; ReaxFF force field validation tests; method used for creation of mixed halide perovskites; molecular dynamics simulation details; methods for extraction and analysis of octahedral tilting in perovskites; analysis of the strain effect and its propagation through the perovskite lattice; and tolerance factors of inorganic halide perovskites
\end{itemize}

\end{suppinfo}


\begin{acknowledgement}

The authors thank Sander Raaijmakers for the useful discussions on the analysis of the perovskite structures. S.T. acknowledges funding from START-UP (Project No. 740.018.024) and Vidi (Project No. VI.Vidi.213.091) from the Dutch Research Council (NWO).

\end{acknowledgement}

\clearpage

\bibliography{ms}

\providecommand{\latin}[1]{#1}
\makeatletter
\providecommand{\doi}
  {\begingroup\let\do\@makeother\dospecials
  \catcode`\{=1 \catcode`\}=2 \doi@aux}
\providecommand{\doi@aux}[1]{\endgroup\texttt{#1}}
\makeatother
\providecommand*\mcitethebibliography{\thebibliography}
\csname @ifundefined\endcsname{endmcitethebibliography}  {\let\endmcitethebibliography\endthebibliography}{}
\begin{mcitethebibliography}{54}
\providecommand*\natexlab[1]{#1}
\providecommand*\mciteSetBstSublistMode[1]{}
\providecommand*\mciteSetBstMaxWidthForm[2]{}
\providecommand*\mciteBstWouldAddEndPuncttrue
  {\def\EndOfBibitem{\unskip.}}
\providecommand*\mciteBstWouldAddEndPunctfalse
  {\let\EndOfBibitem\relax}
\providecommand*\mciteSetBstMidEndSepPunct[3]{}
\providecommand*\mciteSetBstSublistLabelBeginEnd[3]{}
\providecommand*\EndOfBibitem{}
\mciteSetBstSublistMode{f}
\mciteSetBstMaxWidthForm{subitem}{(\alph{mcitesubitemcount})}
\mciteSetBstSublistLabelBeginEnd
  {\mcitemaxwidthsubitemform\space}
  {\relax}
  {\relax}

\bibitem[Green \latin{et~al.}(2014)Green, {Ho-Baillie}, and Snaith]{greenEmergencePerovskiteSolar2014}
Green,~M.~A.; {Ho-Baillie},~A.; Snaith,~H.~J. The Emergence of Perovskite Solar Cells. \emph{Nat. Photonics} \textbf{2014}, \emph{8}, 506--514\relax
\mciteBstWouldAddEndPuncttrue
\mciteSetBstMidEndSepPunct{\mcitedefaultmidpunct}
{\mcitedefaultendpunct}{\mcitedefaultseppunct}\relax
\EndOfBibitem
\bibitem[Liu \latin{et~al.}(2021)Liu, Xu, Bai, Jin, Wang, Friend, and Gao]{liuMetalHalidePerovskites2021}
Liu,~X.-K.; Xu,~W.; Bai,~S.; Jin,~Y.; Wang,~J.; Friend,~R.~H.; Gao,~F. Metal Halide Perovskites for Light-Emitting Diodes. \emph{Nat. Mater.} \textbf{2021}, \emph{20}, 10--21\relax
\mciteBstWouldAddEndPuncttrue
\mciteSetBstMidEndSepPunct{\mcitedefaultmidpunct}
{\mcitedefaultendpunct}{\mcitedefaultseppunct}\relax
\EndOfBibitem
\bibitem[Wang \latin{et~al.}(2021)Wang, Li, Liu, Shi, Fang, and He]{wangLowDimensionalMetalHalide2021}
Wang,~H.-P.; Li,~S.; Liu,~X.; Shi,~Z.; Fang,~X.; He,~J.-H. Low-{{Dimensional Metal Halide Perovskite Photodetectors}}. \emph{Adv. Mater.} \textbf{2021}, \emph{33}, 2003309\relax
\mciteBstWouldAddEndPuncttrue
\mciteSetBstMidEndSepPunct{\mcitedefaultmidpunct}
{\mcitedefaultendpunct}{\mcitedefaultseppunct}\relax
\EndOfBibitem
\bibitem[Wu \latin{et~al.}(2021)Wu, Chen, Yip, and Jen]{wuEvolutionFutureMetal2021}
Wu,~S.; Chen,~Z.; Yip,~H.-L.; Jen,~A. K.~Y. The Evolution and Future of Metal Halide Perovskite-Based Optoelectronic Devices. \emph{Matter} \textbf{2021}, \emph{4}, 3814--3834\relax
\mciteBstWouldAddEndPuncttrue
\mciteSetBstMidEndSepPunct{\mcitedefaultmidpunct}
{\mcitedefaultendpunct}{\mcitedefaultseppunct}\relax
\EndOfBibitem
\bibitem[Nayak \latin{et~al.}(2016)Nayak, Moore, Wenger, Nayak, Haghighirad, Fineberg, Noel, Reid, Rumbles, Kukura, Vincent, and Snaith]{nayakMechanismRapidGrowth2016}
Nayak,~P.~K.; Moore,~D.~T.; Wenger,~B.; Nayak,~S.; Haghighirad,~A.~A.; Fineberg,~A.; Noel,~N.~K.; Reid,~O.~G.; Rumbles,~G.; Kukura,~P. \latin{et~al.}  Mechanism for Rapid Growth of Organic\textendash Inorganic Halide Perovskite Crystals. \emph{Nat. Commun.} \textbf{2016}, \emph{7}, 13303\relax
\mciteBstWouldAddEndPuncttrue
\mciteSetBstMidEndSepPunct{\mcitedefaultmidpunct}
{\mcitedefaultendpunct}{\mcitedefaultseppunct}\relax
\EndOfBibitem
\bibitem[McMeekin \latin{et~al.}(2017)McMeekin, Wang, Rehman, Pulvirenti, Patel, Noel, Johnston, Marder, Herz, and Snaith]{mcmeekinCrystallizationKineticsMorphology2017}
McMeekin,~D.~P.; Wang,~Z.; Rehman,~W.; Pulvirenti,~F.; Patel,~J.~B.; Noel,~N.~K.; Johnston,~M.~B.; Marder,~S.~R.; Herz,~L.~M.; Snaith,~H.~J. Crystallization {{Kinetics}} and {{Morphology Control}} of {{Formamidinium}}\textendash{{Cesium Mixed-Cation Lead Mixed-Halide Perovskite}} via {{Tunability}} of the {{Colloidal Precursor Solution}}. \emph{Adv. Mater.} \textbf{2017}, \emph{29}, 1607039\relax
\mciteBstWouldAddEndPuncttrue
\mciteSetBstMidEndSepPunct{\mcitedefaultmidpunct}
{\mcitedefaultendpunct}{\mcitedefaultseppunct}\relax
\EndOfBibitem
\bibitem[Jacobsson \latin{et~al.}(2016)Jacobsson, {Correa-Baena}, Pazoki, Saliba, Schenk, Gr{\"a}tzel, and Hagfeldt]{jacobssonExplorationCompositionalSpace2016}
Jacobsson,~T.~J.; {Correa-Baena},~J.-P.; Pazoki,~M.; Saliba,~M.; Schenk,~K.; Gr{\"a}tzel,~M.; Hagfeldt,~A. Exploration of the Compositional Space for Mixed Lead Halogen Perovskites for High Efficiency Solar Cells. \emph{Energy Environ. Sci.} \textbf{2016}, \emph{9}, 1706--1724\relax
\mciteBstWouldAddEndPuncttrue
\mciteSetBstMidEndSepPunct{\mcitedefaultmidpunct}
{\mcitedefaultendpunct}{\mcitedefaultseppunct}\relax
\EndOfBibitem
\bibitem[Saliba \latin{et~al.}(2016)Saliba, Matsui, Seo, Domanski, {Correa-Baena}, Nazeeruddin, Zakeeruddin, Tress, Abate, Hagfeldt, and Gr{\"a}tzel]{salibaCesiumcontainingTripleCation2016}
Saliba,~M.; Matsui,~T.; Seo,~J.-Y.; Domanski,~K.; {Correa-Baena},~J.-P.; Nazeeruddin,~M.~K.; Zakeeruddin,~S.~M.; Tress,~W.; Abate,~A.; Hagfeldt,~A. \latin{et~al.}  Cesium-Containing Triple Cation Perovskite Solar Cells: Improved Stability, Reproducibility and High Efficiency. \emph{Energy Environ. Sci.} \textbf{2016}, \emph{9}, 1989--1997\relax
\mciteBstWouldAddEndPuncttrue
\mciteSetBstMidEndSepPunct{\mcitedefaultmidpunct}
{\mcitedefaultendpunct}{\mcitedefaultseppunct}\relax
\EndOfBibitem
\bibitem[Zhu \latin{et~al.}(2018)Zhu, Sun, Zhang, Dai, Xing, Li, Huang, and Huang]{zhuMetalHalidePerovskites2018}
Zhu,~Z.; Sun,~Q.; Zhang,~Z.; Dai,~J.; Xing,~G.; Li,~S.; Huang,~X.; Huang,~W. Metal Halide Perovskites: Stability and Sensing-Ability. \emph{J. Mater. Chem. C} \textbf{2018}, \emph{6}, 10121--10137\relax
\mciteBstWouldAddEndPuncttrue
\mciteSetBstMidEndSepPunct{\mcitedefaultmidpunct}
{\mcitedefaultendpunct}{\mcitedefaultseppunct}\relax
\EndOfBibitem
\bibitem[Kundu and Kelly(2020)Kundu, and Kelly]{kunduSituStudiesDegradation2020}
Kundu,~S.; Kelly,~T.~L. In Situ Studies of the Degradation Mechanisms of Perovskite Solar Cells. \emph{EcoMat} \textbf{2020}, \emph{2}, e12025\relax
\mciteBstWouldAddEndPuncttrue
\mciteSetBstMidEndSepPunct{\mcitedefaultmidpunct}
{\mcitedefaultendpunct}{\mcitedefaultseppunct}\relax
\EndOfBibitem
\bibitem[Conings \latin{et~al.}(2015)Conings, Drijkoningen, Gauquelin, Babayigit, D'Haen, D'Olieslaeger, Ethirajan, Verbeeck, Manca, Mosconi, Angelis, and Boyen]{coningsIntrinsicThermalInstability2015}
Conings,~B.; Drijkoningen,~J.; Gauquelin,~N.; Babayigit,~A.; D'Haen,~J.; D'Olieslaeger,~L.; Ethirajan,~A.; Verbeeck,~J.; Manca,~J.; Mosconi,~E. \latin{et~al.}  Intrinsic {{Thermal Instability}} of {{Methylammonium Lead Trihalide Perovskite}}. \emph{Adv. Energy Mater.} \textbf{2015}, \emph{5}, 1500477\relax
\mciteBstWouldAddEndPuncttrue
\mciteSetBstMidEndSepPunct{\mcitedefaultmidpunct}
{\mcitedefaultendpunct}{\mcitedefaultseppunct}\relax
\EndOfBibitem
\bibitem[Yang \latin{et~al.}(2015)Yang, Siempelkamp, Liu, and Kelly]{yangInvestigationCH3NH3PbI3Degradation2015}
Yang,~J.; Siempelkamp,~B.~D.; Liu,~D.; Kelly,~T.~L. Investigation of {{\ch{CH3NH3PbI3} Degradation Rates}} and {{Mechanisms}} in {{Controlled Humidity Environments Using}} in {{Situ Techniques}}. \emph{ACS Nano} \textbf{2015}, \emph{9}, 1955--1963\relax
\mciteBstWouldAddEndPuncttrue
\mciteSetBstMidEndSepPunct{\mcitedefaultmidpunct}
{\mcitedefaultendpunct}{\mcitedefaultseppunct}\relax
\EndOfBibitem
\bibitem[{Juarez-Perez} \latin{et~al.}(2019){Juarez-Perez}, Ono, and Qi]{juarez-perezThermalDegradationFormamidinium2019}
{Juarez-Perez},~E.~J.; Ono,~L.~K.; Qi,~Y. Thermal Degradation of Formamidinium Based Lead Halide Perovskites into Sym-Triazine and Hydrogen Cyanide Observed by Coupled Thermogravimetry-Mass Spectrometry Analysis. \emph{J. Mater. Chem. A} \textbf{2019}, \emph{7}, 16912--16919\relax
\mciteBstWouldAddEndPuncttrue
\mciteSetBstMidEndSepPunct{\mcitedefaultmidpunct}
{\mcitedefaultendpunct}{\mcitedefaultseppunct}\relax
\EndOfBibitem
\bibitem[Xiang \latin{et~al.}(2021)Xiang, Liu, and Tress]{xiangReviewStabilityInorganic2021}
Xiang,~W.; Liu,~S.~F.; Tress,~W. A Review on the Stability of Inorganic Metal Halide Perovskites: Challenges and Opportunities for Stable Solar Cells. \emph{Energy Environ. Sci.} \textbf{2021}, \emph{14}, 2090--2113\relax
\mciteBstWouldAddEndPuncttrue
\mciteSetBstMidEndSepPunct{\mcitedefaultmidpunct}
{\mcitedefaultendpunct}{\mcitedefaultseppunct}\relax
\EndOfBibitem
\bibitem[Eperon \latin{et~al.}(2015)Eperon, Patern{\`o}, Sutton, Zampetti, Haghighirad, Cacialli, and Snaith]{eperonInorganicCaesiumLead2015}
Eperon,~G.~E.; Patern{\`o},~G.~M.; Sutton,~R.~J.; Zampetti,~A.; Haghighirad,~A.~A.; Cacialli,~F.; Snaith,~H.~J. Inorganic Caesium Lead Iodide Perovskite Solar Cells. \emph{J. Mater. Chem. A} \textbf{2015}, \emph{3}, 19688--19695\relax
\mciteBstWouldAddEndPuncttrue
\mciteSetBstMidEndSepPunct{\mcitedefaultmidpunct}
{\mcitedefaultendpunct}{\mcitedefaultseppunct}\relax
\EndOfBibitem
\bibitem[Duan \latin{et~al.}(2022)Duan, Zhang, Liu, Gr{\"a}tzel, and Luo]{duanPhasePureGCsPbI3Efficient2022}
Duan,~L.; Zhang,~H.; Liu,~M.; Gr{\"a}tzel,~M.; Luo,~J. Phase-{{Pure}} {\textgamma}-{{\ch{CsPbI3}}} for {{Efficient Inorganic Perovskite Solar Cells}}. \emph{ACS Energy Lett.} \textbf{2022}, \emph{7}, 2911--2918\relax
\mciteBstWouldAddEndPuncttrue
\mciteSetBstMidEndSepPunct{\mcitedefaultmidpunct}
{\mcitedefaultendpunct}{\mcitedefaultseppunct}\relax
\EndOfBibitem
\bibitem[Chen \latin{et~al.}(2023)Chen, Ko, Li, Zou, Geng, Guo, and Halpert]{chenAminoAcidPassivatedPure2023}
Chen,~D.; Ko,~P.~K.; Li,~C.-H.~A.; Zou,~B.; Geng,~P.; Guo,~L.; Halpert,~J.~E. Amino {{Acid-Passivated Pure Red \ch{CsPbI3} Quantum Dot LEDs}}. \emph{ACS Energy Lett.} \textbf{2023}, \emph{8}, 410--416\relax
\mciteBstWouldAddEndPuncttrue
\mciteSetBstMidEndSepPunct{\mcitedefaultmidpunct}
{\mcitedefaultendpunct}{\mcitedefaultseppunct}\relax
\EndOfBibitem
\bibitem[Mannino \latin{et~al.}(2020)Mannino, Deretzis, Smecca, La~Magna, Alberti, Ceratti, and Cahen]{manninoTemperatureDependentOpticalBand2020}
Mannino,~G.; Deretzis,~I.; Smecca,~E.; La~Magna,~A.; Alberti,~A.; Ceratti,~D.; Cahen,~D. Temperature-{{Dependent Optical Band Gap}} in {{\ch{CsPbBr3}}}, {{\ch{MAPbBr3}}}, and {{\ch{FAPbBr3} Single Crystals}}. \emph{J. Phys. Chem. Lett.} \textbf{2020}, \emph{11}, 2490--2496\relax
\mciteBstWouldAddEndPuncttrue
\mciteSetBstMidEndSepPunct{\mcitedefaultmidpunct}
{\mcitedefaultendpunct}{\mcitedefaultseppunct}\relax
\EndOfBibitem
\bibitem[Li \latin{et~al.}(2019)Li, Tan, Lai, Li, Chen, Li, Xu, and Yang]{liAllInorganicCsPbBr3Perovskite2019}
Li,~X.; Tan,~Y.; Lai,~H.; Li,~S.; Chen,~Y.; Li,~S.; Xu,~P.; Yang,~J. All-{{Inorganic \ch{CsPbBr3} Perovskite Solar Cells}} with 10.45\% {{Efficiency}} by {{Evaporation-Assisted Deposition}} and {{Setting Intermediate Energy Levels}}. \emph{ACS Appl. Mater. Interfaces} \textbf{2019}, \emph{11}, 29746--29752\relax
\mciteBstWouldAddEndPuncttrue
\mciteSetBstMidEndSepPunct{\mcitedefaultmidpunct}
{\mcitedefaultendpunct}{\mcitedefaultseppunct}\relax
\EndOfBibitem
\bibitem[Wang \latin{et~al.}(2019)Wang, Zhang, Wu, Cao, Yang, Shang, Ning, Zhang, Zheng, Yan, Kershaw, Zhang, Rogach, and Yang]{wangTrifluoroacetateInducedSmallgrained2019}
Wang,~H.; Zhang,~X.; Wu,~Q.; Cao,~F.; Yang,~D.; Shang,~Y.; Ning,~Z.; Zhang,~W.; Zheng,~W.; Yan,~Y. \latin{et~al.}  Trifluoroacetate Induced Small-Grained {{\ch{CsPbBr3}}} Perovskite Films Result in Efficient and Stable Light-Emitting Devices. \emph{Nat. Commun.} \textbf{2019}, \emph{10}, 665\relax
\mciteBstWouldAddEndPuncttrue
\mciteSetBstMidEndSepPunct{\mcitedefaultmidpunct}
{\mcitedefaultendpunct}{\mcitedefaultseppunct}\relax
\EndOfBibitem
\bibitem[Stoumpos \latin{et~al.}(2013)Stoumpos, Malliakas, Peters, Liu, Sebastian, Im, Chasapis, Wibowo, Chung, Freeman, Wessels, and Kanatzidis]{stoumposCrystalGrowthPerovskite2013}
Stoumpos,~C.~C.; Malliakas,~C.~D.; Peters,~J.~A.; Liu,~Z.; Sebastian,~M.; Im,~J.; Chasapis,~T.~C.; Wibowo,~A.~C.; Chung,~D.~Y.; Freeman,~A.~J. \latin{et~al.}  Crystal {{Growth}} of the {{Perovskite Semiconductor \ch{CsPbBr3}}}: {{A New Material}} for {{High-Energy Radiation Detection}}. \emph{Cryst. Growth Des.} \textbf{2013}, \emph{13}, 2722--2727\relax
\mciteBstWouldAddEndPuncttrue
\mciteSetBstMidEndSepPunct{\mcitedefaultmidpunct}
{\mcitedefaultendpunct}{\mcitedefaultseppunct}\relax
\EndOfBibitem
\bibitem[Protesescu \latin{et~al.}(2015)Protesescu, Yakunin, Bodnarchuk, Krieg, Caputo, Hendon, Yang, Walsh, and Kovalenko]{protesescuNanocrystalsCesiumLead2015}
Protesescu,~L.; Yakunin,~S.; Bodnarchuk,~M.~I.; Krieg,~F.; Caputo,~R.; Hendon,~C.~H.; Yang,~R.~X.; Walsh,~A.; Kovalenko,~M.~V. Nanocrystals of {{Cesium Lead Halide Perovskites}} ({{\ch{CsPbX3}}}, {{X}} = {\ch{Cl}}, {\ch{Br}}, and {\ch{I}}): {{Novel Optoelectronic Materials Showing Bright Emission}} with {{Wide Color Gamut}}. \emph{Nano Lett.} \textbf{2015}, \emph{15}, 3692--3696\relax
\mciteBstWouldAddEndPuncttrue
\mciteSetBstMidEndSepPunct{\mcitedefaultmidpunct}
{\mcitedefaultendpunct}{\mcitedefaultseppunct}\relax
\EndOfBibitem
\bibitem[Stoumpos and Kanatzidis(2015)Stoumpos, and Kanatzidis]{stoumposRenaissanceHalidePerovskites2015}
Stoumpos,~C.~C.; Kanatzidis,~M.~G. The {{Renaissance}} of {{Halide Perovskites}} and {{Their Evolution}} as {{Emerging Semiconductors}}. \emph{Acc. Chem. Res.} \textbf{2015}, \emph{48}, 2791--2802\relax
\mciteBstWouldAddEndPuncttrue
\mciteSetBstMidEndSepPunct{\mcitedefaultmidpunct}
{\mcitedefaultendpunct}{\mcitedefaultseppunct}\relax
\EndOfBibitem
\bibitem[Marronnier \latin{et~al.}(2018)Marronnier, Roma, {Boyer-Richard}, Pedesseau, Jancu, Bonnassieux, Katan, Stoumpos, Kanatzidis, and Even]{marronnierAnharmonicityDisorderBlack2018}
Marronnier,~A.; Roma,~G.; {Boyer-Richard},~S.; Pedesseau,~L.; Jancu,~J.-M.; Bonnassieux,~Y.; Katan,~C.; Stoumpos,~C.~C.; Kanatzidis,~M.~G.; Even,~J. Anharmonicity and {{Disorder}} in the {{Black Phases}} of {{Cesium Lead Iodide Used}} for {{Stable Inorganic Perovskite Solar Cells}}. \emph{ACS Nano} \textbf{2018}, \emph{12}, 3477--3486\relax
\mciteBstWouldAddEndPuncttrue
\mciteSetBstMidEndSepPunct{\mcitedefaultmidpunct}
{\mcitedefaultendpunct}{\mcitedefaultseppunct}\relax
\EndOfBibitem
\bibitem[Goldschmidt(1926)]{goldschmidtGesetzeKrystallochemie1926}
Goldschmidt,~V.~M. {Die Gesetze der Krystallochemie}. \emph{Naturwissenschaften} \textbf{1926}, \emph{14}, 477--485\relax
\mciteBstWouldAddEndPuncttrue
\mciteSetBstMidEndSepPunct{\mcitedefaultmidpunct}
{\mcitedefaultendpunct}{\mcitedefaultseppunct}\relax
\EndOfBibitem
\bibitem[Shannon(1976)]{shannonRevisedEffectiveIonic1976}
Shannon,~R.~D. Revised Effective Ionic Radii and Systematic Studies of Interatomic Distances in Halides and Chalcogenides. \emph{Acta Crystallogr. A.} \textbf{1976}, \emph{32}, 751--767\relax
\mciteBstWouldAddEndPuncttrue
\mciteSetBstMidEndSepPunct{\mcitedefaultmidpunct}
{\mcitedefaultendpunct}{\mcitedefaultseppunct}\relax
\EndOfBibitem
\bibitem[Straus \latin{et~al.}(2020)Straus, Guo, Abeykoon, and Cava]{strausUnderstandingInstabilityHalide2020}
Straus,~D.~B.; Guo,~S.; Abeykoon,~A.~M.; Cava,~R.~J. Understanding the {{Instability}} of the {{Halide Perovskite \ch{CsPbI3}}} through {{Temperature-Dependent Structural Analysis}}. \emph{Adv. Mater.} \textbf{2020}, \emph{32}, 2001069\relax
\mciteBstWouldAddEndPuncttrue
\mciteSetBstMidEndSepPunct{\mcitedefaultmidpunct}
{\mcitedefaultendpunct}{\mcitedefaultseppunct}\relax
\EndOfBibitem
\bibitem[Beal \latin{et~al.}(2016)Beal, Slotcavage, Leijtens, Bowring, Belisle, Nguyen, Burkhard, Hoke, and McGehee]{bealCesiumLeadHalide2016}
Beal,~R.~E.; Slotcavage,~D.~J.; Leijtens,~T.; Bowring,~A.~R.; Belisle,~R.~A.; Nguyen,~W.~H.; Burkhard,~G.~F.; Hoke,~E.~T.; McGehee,~M.~D. Cesium {{Lead Halide Perovskites}} with {{Improved Stability}} for {{Tandem Solar Cells}}. \emph{J. Phys. Chem. Lett.} \textbf{2016}, \emph{7}, 746--751\relax
\mciteBstWouldAddEndPuncttrue
\mciteSetBstMidEndSepPunct{\mcitedefaultmidpunct}
{\mcitedefaultendpunct}{\mcitedefaultseppunct}\relax
\EndOfBibitem
\bibitem[Sutton \latin{et~al.}(2016)Sutton, Eperon, Miranda, Parrott, Kamino, Patel, H{\"o}rantner, Johnston, Haghighirad, Moore, and Snaith]{suttonBandgapTunableCesiumLead2016}
Sutton,~R.~J.; Eperon,~G.~E.; Miranda,~L.; Parrott,~E.~S.; Kamino,~B.~A.; Patel,~J.~B.; H{\"o}rantner,~M.~T.; Johnston,~M.~B.; Haghighirad,~A.~A.; Moore,~D.~T. \latin{et~al.}  Bandgap-{{Tunable Cesium Lead Halide Perovskites}} with {{High Thermal Stability}} for {{Efficient Solar Cells}}. \emph{Adv. Energy Mater.} \textbf{2016}, \emph{6}, 1502458\relax
\mciteBstWouldAddEndPuncttrue
\mciteSetBstMidEndSepPunct{\mcitedefaultmidpunct}
{\mcitedefaultendpunct}{\mcitedefaultseppunct}\relax
\EndOfBibitem
\bibitem[Ghosh \latin{et~al.}(2018)Ghosh, Ali, Chaudhary, and Bhattacharyya]{ghoshDependenceHalideComposition2018}
Ghosh,~D.; Ali,~M.~Y.; Chaudhary,~D.~K.; Bhattacharyya,~S. Dependence of Halide Composition on the Stability of Highly Efficient All-Inorganic Cesium Lead Halide Perovskite Quantum Dot Solar Cells. \emph{Sol. Energy Mater. Sol. Cells} \textbf{2018}, \emph{185}, 28--35\relax
\mciteBstWouldAddEndPuncttrue
\mciteSetBstMidEndSepPunct{\mcitedefaultmidpunct}
{\mcitedefaultendpunct}{\mcitedefaultseppunct}\relax
\EndOfBibitem
\bibitem[Chen \latin{et~al.}(2017)Chen, Lin, Chiang, Tsai, Huang, Tsao, and Lin]{chenAllVacuumDepositedStoichiometricallyBalanced2017}
Chen,~C.-Y.; Lin,~H.-Y.; Chiang,~K.-M.; Tsai,~W.-L.; Huang,~Y.-C.; Tsao,~C.-S.; Lin,~H.-W. All-{{Vacuum-Deposited Stoichiometrically Balanced Inorganic Cesium Lead Halide Perovskite Solar Cells}} with {{Stabilized Efficiency Exceeding}} 11\%. \emph{Adv. Mater.} \textbf{2017}, \emph{29}, 1605290\relax
\mciteBstWouldAddEndPuncttrue
\mciteSetBstMidEndSepPunct{\mcitedefaultmidpunct}
{\mcitedefaultendpunct}{\mcitedefaultseppunct}\relax
\EndOfBibitem
\bibitem[Li \latin{et~al.}(2018)Li, Xu, Zhou, Zhang, Liu, Dai, and Yao]{liCsBrInducedStableCsPbI32018}
Li,~Z.; Xu,~J.; Zhou,~S.; Zhang,~B.; Liu,~X.; Dai,~S.; Yao,~J. {{CsBr-Induced Stable \ch{CsPbI_{3-x}Br_{x}}}} (x {$<$} 1) {{Perovskite Films}} at {{Low Temperature}} for {{Highly Efficient Planar Heterojunction Solar Cells}}. \emph{ACS Appl. Mater. Interfaces} \textbf{2018}, \emph{10}, 38183--38192\relax
\mciteBstWouldAddEndPuncttrue
\mciteSetBstMidEndSepPunct{\mcitedefaultmidpunct}
{\mcitedefaultendpunct}{\mcitedefaultseppunct}\relax
\EndOfBibitem
\bibitem[Lin \latin{et~al.}(2018)Lin, Lai, Dou, Kley, Chen, Peng, Sun, Lu, Hawks, Xie, Cui, Alivisatos, Limmer, and Yang]{linThermochromicHalidePerovskite2018}
Lin,~J.; Lai,~M.; Dou,~L.; Kley,~C.~S.; Chen,~H.; Peng,~F.; Sun,~J.; Lu,~D.; Hawks,~S.~A.; Xie,~C. \latin{et~al.}  Thermochromic Halide Perovskite Solar Cells. \emph{Nat. Mater.} \textbf{2018}, \emph{17}, 261--267\relax
\mciteBstWouldAddEndPuncttrue
\mciteSetBstMidEndSepPunct{\mcitedefaultmidpunct}
{\mcitedefaultendpunct}{\mcitedefaultseppunct}\relax
\EndOfBibitem
\bibitem[Liu \latin{et~al.}(2021)Liu, Li, Zhang, Yang, Zhao, and Chen]{liuCsPbI3PerovskiteQuantum2021}
Liu,~Y.; Li,~Q.; Zhang,~W.; Yang,~Z.; Zhao,~S.; Chen,~W. {{\ch{CsPbI3} Perovskite Quantum Dot Solar Cells}} with {{Both High Efficiency}} and {{Phase Stability Enabled}} by {{Br Doping}}. \emph{ACS Appl. Energy Mater.} \textbf{2021}, \emph{4}, 6688--6696\relax
\mciteBstWouldAddEndPuncttrue
\mciteSetBstMidEndSepPunct{\mcitedefaultmidpunct}
{\mcitedefaultendpunct}{\mcitedefaultseppunct}\relax
\EndOfBibitem
\bibitem[N{\"a}sstr{\"o}m \latin{et~al.}(2020)N{\"a}sstr{\"o}m, Becker, M{\'a}rquez, Shargaieva, Mainz, Unger, and Unold]{nasstromDependencePhaseTransitions2020}
N{\"a}sstr{\"o}m,~H.; Becker,~P.; M{\'a}rquez,~J.~A.; Shargaieva,~O.; Mainz,~R.; Unger,~E.; Unold,~T. Dependence of Phase Transitions on Halide Ratio in Inorganic \ch{CsPb(Br_{x}I_{1-x})3} Perovskite Thin Films Obtained from High-Throughput Experimentation. \emph{J. Mater. Chem. A} \textbf{2020}, \emph{8}, 22626--22631\relax
\mciteBstWouldAddEndPuncttrue
\mciteSetBstMidEndSepPunct{\mcitedefaultmidpunct}
{\mcitedefaultendpunct}{\mcitedefaultseppunct}\relax
\EndOfBibitem
\bibitem[Pols \latin{et~al.}(2021)Pols, {Vicent-Luna}, Filot, {van Duin}, and Tao]{polsAtomisticInsightsDegradation2021}
Pols,~M.; {Vicent-Luna},~J.~M.; Filot,~I.; {van Duin},~A. C.~T.; Tao,~S. Atomistic {{Insights Into}} the {{Degradation}} of {Inorganic Halide Perovskite \ch{CsPbI3}}: {{A Reactive Force Field Molecular Dynamics Study}}. \emph{J. Phys. Chem. Lett.} \textbf{2021}, \emph{12}, 5519--5525\relax
\mciteBstWouldAddEndPuncttrue
\mciteSetBstMidEndSepPunct{\mcitedefaultmidpunct}
{\mcitedefaultendpunct}{\mcitedefaultseppunct}\relax
\EndOfBibitem
\bibitem[Pols \latin{et~al.}(2022)Pols, Hilpert, Filot, {van Duin}, Calero, and Tao]{polsWhatHappensSurfaces2022}
Pols,~M.; Hilpert,~T.; Filot,~I.~A.; {van Duin},~A.~C.; Calero,~S.; Tao,~S. What {{Happens}} at {{Surfaces}} and {{Grain Boundaries}} of {{Halide Perovskites}}: {{Insights}} from {{Reactive Molecular Dynamics Simulations}} of {\ch{CsPbI3}}. \emph{ACS Appl. Mater. Interfaces} \textbf{2022}, \emph{14}, 40841--40850\relax
\mciteBstWouldAddEndPuncttrue
\mciteSetBstMidEndSepPunct{\mcitedefaultmidpunct}
{\mcitedefaultendpunct}{\mcitedefaultseppunct}\relax
\EndOfBibitem
\bibitem[Kresse and Hafner(1994)Kresse, and Hafner]{kresseInitioMoleculardynamicsSimulation1994}
Kresse,~G.; Hafner,~J. Ab Initio Molecular-Dynamics Simulation of the Liquid-Metal--Amorphous-Semiconductor Transition in Germanium. \emph{Phys. Rev. B} \textbf{1994}, \emph{49}, 14251--14269\relax
\mciteBstWouldAddEndPuncttrue
\mciteSetBstMidEndSepPunct{\mcitedefaultmidpunct}
{\mcitedefaultendpunct}{\mcitedefaultseppunct}\relax
\EndOfBibitem
\bibitem[Kresse and Furthm{\"u}ller(1996)Kresse, and Furthm{\"u}ller]{kresseEfficiencyAbinitioTotal1996}
Kresse,~G.; Furthm{\"u}ller,~J. Efficiency of Ab-Initio Total Energy Calculations for Metals and Semiconductors Using a Plane-Wave Basis Set. \emph{Comput. Mater. Sci.} \textbf{1996}, \emph{6}, 15--50\relax
\mciteBstWouldAddEndPuncttrue
\mciteSetBstMidEndSepPunct{\mcitedefaultmidpunct}
{\mcitedefaultendpunct}{\mcitedefaultseppunct}\relax
\EndOfBibitem
\bibitem[Kresse and Furthm{\"u}ller(1996)Kresse, and Furthm{\"u}ller]{kresseEfficientIterativeSchemes1996}
Kresse,~G.; Furthm{\"u}ller,~J. Efficient Iterative Schemes for Ab Initio Total-Energy Calculations Using a Plane-Wave Basis Set. \emph{Phys. Rev. B} \textbf{1996}, \emph{54}, 11169--11186\relax
\mciteBstWouldAddEndPuncttrue
\mciteSetBstMidEndSepPunct{\mcitedefaultmidpunct}
{\mcitedefaultendpunct}{\mcitedefaultseppunct}\relax
\EndOfBibitem
\bibitem[Fonseca~Guerra \latin{et~al.}(1998)Fonseca~Guerra, Snijders, {te Velde}, and Baerends]{fonsecaguerraOrderNDFTMethod1998}
Fonseca~Guerra,~C.; Snijders,~J.~G.; {te Velde},~G.; Baerends,~E.~J. Towards an Order-{{N DFT}} Method. \emph{Theor. Chem. Acc.} \textbf{1998}, \emph{99}, 391--403\relax
\mciteBstWouldAddEndPuncttrue
\mciteSetBstMidEndSepPunct{\mcitedefaultmidpunct}
{\mcitedefaultendpunct}{\mcitedefaultseppunct}\relax
\EndOfBibitem
\bibitem[{te Velde} \latin{et~al.}(2001){te Velde}, Bickelhaupt, Baerends, Fonseca~Guerra, {van Gisbergen}, Snijders, and Ziegler]{teveldeChemistryADF2001}
{te Velde},~G.; Bickelhaupt,~F.~M.; Baerends,~E.~J.; Fonseca~Guerra,~C.; {van Gisbergen},~S. J.~A.; Snijders,~J.~G.; Ziegler,~T. Chemistry with {{ADF}}. \emph{J. Comput. Chem.} \textbf{2001}, \emph{22}, 931--967\relax
\mciteBstWouldAddEndPuncttrue
\mciteSetBstMidEndSepPunct{\mcitedefaultmidpunct}
{\mcitedefaultendpunct}{\mcitedefaultseppunct}\relax
\EndOfBibitem
\bibitem[Perdew \latin{et~al.}(1996)Perdew, Burke, and Ernzerhof]{perdewGeneralizedGradientApproximation1996}
Perdew,~J.~P.; Burke,~K.; Ernzerhof,~M. Generalized {{Gradient Approximation Made Simple}}. \emph{Phys. Rev. Lett.} \textbf{1996}, \emph{77}, 3865--3868\relax
\mciteBstWouldAddEndPuncttrue
\mciteSetBstMidEndSepPunct{\mcitedefaultmidpunct}
{\mcitedefaultendpunct}{\mcitedefaultseppunct}\relax
\EndOfBibitem
\bibitem[Grimme \latin{et~al.}(2011)Grimme, Ehrlich, and Goerigk]{grimmeEffectDampingFunction2011}
Grimme,~S.; Ehrlich,~S.; Goerigk,~L. Effect of the Damping Function in Dispersion Corrected Density Functional Theory. \emph{J. Comput. Chem.} \textbf{2011}, \emph{32}, 1456--1465\relax
\mciteBstWouldAddEndPuncttrue
\mciteSetBstMidEndSepPunct{\mcitedefaultmidpunct}
{\mcitedefaultendpunct}{\mcitedefaultseppunct}\relax
\EndOfBibitem
\bibitem[Hansen and Ostermeier(2001)Hansen, and Ostermeier]{hansenCompletelyDerandomizedSelfAdaptation2001}
Hansen,~N.; Ostermeier,~A. Completely {{Derandomized Self-Adaptation}} in {{Evolution Strategies}}. \emph{Evol. Comput.} \textbf{2001}, \emph{9}, 159--195\relax
\mciteBstWouldAddEndPuncttrue
\mciteSetBstMidEndSepPunct{\mcitedefaultmidpunct}
{\mcitedefaultendpunct}{\mcitedefaultseppunct}\relax
\EndOfBibitem
\bibitem[Komissarov \latin{et~al.}(2021)Komissarov, R{\"u}ger, Hellstr{\"o}m, and Verstraelen]{komissarovParAMSParameterOptimization2021}
Komissarov,~L.; R{\"u}ger,~R.; Hellstr{\"o}m,~M.; Verstraelen,~T. {{ParAMS}}: {{Parameter Optimization}} for {{Atomistic}} and {{Molecular Simulations}}. \emph{J. Chem. Inf. Model.} \textbf{2021}, \emph{61}, 3737--3743\relax
\mciteBstWouldAddEndPuncttrue
\mciteSetBstMidEndSepPunct{\mcitedefaultmidpunct}
{\mcitedefaultendpunct}{\mcitedefaultseppunct}\relax
\EndOfBibitem
\bibitem[AMS()]{AMS2022}
{Rüger, R.; Franchini, M.; Trnka, T.; Yakovlev, A.; Philipsen, P.; van Vuren, T., Klumpers, B., Soini, T.}, \textit{AMS 2022}, {SCM, Theoretical Chemistry, Vrije Universiteit, Amsterdam, The Netherlands}, {2022}\relax
\mciteBstWouldAddEndPuncttrue
\mciteSetBstMidEndSepPunct{\mcitedefaultmidpunct}
{\mcitedefaultendpunct}{\mcitedefaultseppunct}\relax
\EndOfBibitem
\bibitem[Chen \latin{et~al.}(2022)Chen, Guo, Gong, and Yin]{chenKineticPathwayGtod2022}
Chen,~G.-Y.; Guo,~Z.-D.; Gong,~X.-G.; Yin,~W.-J. Kinetic Pathway of {\textgamma}-to-{\textdelta} Phase Transition in {\ch{CsPbI3}}. \emph{Chem} \textbf{2022}, \emph{8}, 3120--3129\relax
\mciteBstWouldAddEndPuncttrue
\mciteSetBstMidEndSepPunct{\mcitedefaultmidpunct}
{\mcitedefaultendpunct}{\mcitedefaultseppunct}\relax
\EndOfBibitem
\bibitem[Yang \latin{et~al.}(2017)Yang, Skelton, {da Silva}, Frost, and Walsh]{yangSpontaneousOctahedralTilting2017}
Yang,~R.~X.; Skelton,~J.~M.; {da Silva},~E.~L.; Frost,~J.~M.; Walsh,~A. Spontaneous {{Octahedral Tilting}} in the {{Cubic Inorganic Cesium Halide Perovskites \ch{CsSnX3}}} and {\ch{CsPbX3}} ({\ch{X}} = {\ch{F}}, {\ch{Cl}}, {\ch{Br}}, {\ch{I}}). \emph{J. Phys. Chem. Lett.} \textbf{2017}, \emph{8}, 4720--4726\relax
\mciteBstWouldAddEndPuncttrue
\mciteSetBstMidEndSepPunct{\mcitedefaultmidpunct}
{\mcitedefaultendpunct}{\mcitedefaultseppunct}\relax
\EndOfBibitem
\bibitem[Jinnouchi \latin{et~al.}(2019)Jinnouchi, Lahnsteiner, Karsai, Kresse, and Bokdam]{jinnouchiPhaseTransitionsHybrid2019}
Jinnouchi,~R.; Lahnsteiner,~J.; Karsai,~F.; Kresse,~G.; Bokdam,~M. Phase {{Transitions}} of {{Hybrid Perovskites Simulated}} by {{Machine-Learning Force Fields Trained}} on the {{Fly}} with {{Bayesian Inference}}. \emph{Phys. Rev. Lett.} \textbf{2019}, \emph{122}, 225701\relax
\mciteBstWouldAddEndPuncttrue
\mciteSetBstMidEndSepPunct{\mcitedefaultmidpunct}
{\mcitedefaultendpunct}{\mcitedefaultseppunct}\relax
\EndOfBibitem
\bibitem[Fransson \latin{et~al.}(2023)Fransson, Wiktor, and Erhart]{franssonPhaseTransitionsInorganic2023}
Fransson,~E.; Wiktor,~J.; Erhart,~P. Phase {{Transitions}} in {{Inorganic Halide Perovskites}} from {{Machine-Learned Potentials}}. \emph{J. Phys. Chem. C} \textbf{2023}, \emph{127}, 13773--13781\relax
\mciteBstWouldAddEndPuncttrue
\mciteSetBstMidEndSepPunct{\mcitedefaultmidpunct}
{\mcitedefaultendpunct}{\mcitedefaultseppunct}\relax
\EndOfBibitem
\bibitem[Sharma \latin{et~al.}(1992)Sharma, Weiden, and Weiss]{sharmaPhaseDiagramsQuasibinary1992}
Sharma,~S.; Weiden,~N.; Weiss,~A. Phase {{Diagrams}} of {{Quasibinary Systems}} of the {{Type}}: {\ch{ABX3}} -- {\ch{A'BX3}}; {\ch{ABX3}} -- {\ch{AB'X3}}; {\ch{ABX3}} -- {\ch{ABX'3}}; {\ch{X}} = {{Halogen}}. \emph{Z. Phys. Chem.} \textbf{1992}, \emph{175}, 63--80\relax
\mciteBstWouldAddEndPuncttrue
\mciteSetBstMidEndSepPunct{\mcitedefaultmidpunct}
{\mcitedefaultendpunct}{\mcitedefaultseppunct}\relax
\EndOfBibitem
\bibitem[Wiktor \latin{et~al.}(2023)Wiktor, Fransson, Kubicki, and Erhart]{wiktorQuantifyingDynamicTilting2023}
Wiktor,~J.; Fransson,~E.; Kubicki,~D.; Erhart,~P. Quantifying {{Dynamic Tilting}} in {{Halide Perovskites}}: {{Chemical Trends}} and {{Local Correlations}}. \emph{Chem. Mater.} \textbf{2023}, \emph{35}, 6737--6744\relax
\mciteBstWouldAddEndPuncttrue
\mciteSetBstMidEndSepPunct{\mcitedefaultmidpunct}
{\mcitedefaultendpunct}{\mcitedefaultseppunct}\relax
\EndOfBibitem
\end{mcitethebibliography}


\providecommand{\latin}[1]{#1}
\makeatletter
\providecommand{\doi}
  {\begingroup\let\do\@makeother\dospecials
  \catcode`\{=1 \catcode`\}=2 \doi@aux}
\providecommand{\doi@aux}[1]{\endgroup\texttt{#1}}
\makeatother
\providecommand*\mcitethebibliography{\thebibliography}
\csname @ifundefined\endcsname{endmcitethebibliography}  {\let\endmcitethebibliography\endthebibliography}{}
\begin{mcitethebibliography}{34}
\providecommand*\natexlab[1]{#1}
\providecommand*\mciteSetBstSublistMode[1]{}
\providecommand*\mciteSetBstMaxWidthForm[2]{}
\providecommand*\mciteBstWouldAddEndPuncttrue
  {\def\EndOfBibitem{\unskip.}}
\providecommand*\mciteBstWouldAddEndPunctfalse
  {\let\EndOfBibitem\relax}
\providecommand*\mciteSetBstMidEndSepPunct[3]{}
\providecommand*\mciteSetBstSublistLabelBeginEnd[3]{}
\providecommand*\EndOfBibitem{}
\mciteSetBstSublistMode{f}
\mciteSetBstMaxWidthForm{subitem}{(\alph{mcitesubitemcount})}
\mciteSetBstSublistLabelBeginEnd
  {\mcitemaxwidthsubitemform\space}
  {\relax}
  {\relax}

\bibitem[Bl{\"o}chl(1994)]{blochlProjectorAugmentedwaveMethod1994}
Bl{\"o}chl,~P.~E. Projector Augmented-Wave Method. \emph{Phys. Rev. B} \textbf{1994}, \emph{50}, 17953--17979\relax
\mciteBstWouldAddEndPuncttrue
\mciteSetBstMidEndSepPunct{\mcitedefaultmidpunct}
{\mcitedefaultendpunct}{\mcitedefaultseppunct}\relax
\EndOfBibitem
\bibitem[Kresse and Joubert(1999)Kresse, and Joubert]{kresseUltrasoftPseudopotentialsProjector1999}
Kresse,~G.; Joubert,~D. From Ultrasoft Pseudopotentials to the Projector Augmented-Wave Method. \emph{Phys. Rev. B} \textbf{1999}, \emph{59}, 1758--1775\relax
\mciteBstWouldAddEndPuncttrue
\mciteSetBstMidEndSepPunct{\mcitedefaultmidpunct}
{\mcitedefaultendpunct}{\mcitedefaultseppunct}\relax
\EndOfBibitem
\bibitem[Kresse and Hafner(1994)Kresse, and Hafner]{kresseInitioMoleculardynamicsSimulation1994}
Kresse,~G.; Hafner,~J. Ab Initio Molecular-Dynamics Simulation of the Liquid-Metal--Amorphous-Semiconductor Transition in Germanium. \emph{Phys. Rev. B} \textbf{1994}, \emph{49}, 14251--14269\relax
\mciteBstWouldAddEndPuncttrue
\mciteSetBstMidEndSepPunct{\mcitedefaultmidpunct}
{\mcitedefaultendpunct}{\mcitedefaultseppunct}\relax
\EndOfBibitem
\bibitem[Kresse and Furthm{\"u}ller(1996)Kresse, and Furthm{\"u}ller]{kresseEfficiencyAbinitioTotal1996}
Kresse,~G.; Furthm{\"u}ller,~J. Efficiency of Ab-Initio Total Energy Calculations for Metals and Semiconductors Using a Plane-Wave Basis Set. \emph{Comput. Mater. Sci.} \textbf{1996}, \emph{6}, 15--50\relax
\mciteBstWouldAddEndPuncttrue
\mciteSetBstMidEndSepPunct{\mcitedefaultmidpunct}
{\mcitedefaultendpunct}{\mcitedefaultseppunct}\relax
\EndOfBibitem
\bibitem[Kresse and Furthm{\"u}ller(1996)Kresse, and Furthm{\"u}ller]{kresseEfficientIterativeSchemes1996}
Kresse,~G.; Furthm{\"u}ller,~J. Efficient Iterative Schemes for Ab Initio Total-Energy Calculations Using a Plane-Wave Basis Set. \emph{Phys. Rev. B} \textbf{1996}, \emph{54}, 11169--11186\relax
\mciteBstWouldAddEndPuncttrue
\mciteSetBstMidEndSepPunct{\mcitedefaultmidpunct}
{\mcitedefaultendpunct}{\mcitedefaultseppunct}\relax
\EndOfBibitem
\bibitem[Pols \latin{et~al.}(2021)Pols, {Vicent-Luna}, Filot, {van Duin}, and Tao]{polsAtomisticInsightsDegradation2021}
Pols,~M.; {Vicent-Luna},~J.~M.; Filot,~I.; {van Duin},~A. C.~T.; Tao,~S. Atomistic {{Insights Into}} the {{Degradation}} of {Inorganic Halide Perovskite \ch{CsPbI3}}: {{A Reactive Force Field Molecular Dynamics Study}}. \emph{J. Phys. Chem. Lett.} \textbf{2021}, \emph{12}, 5519--5525\relax
\mciteBstWouldAddEndPuncttrue
\mciteSetBstMidEndSepPunct{\mcitedefaultmidpunct}
{\mcitedefaultendpunct}{\mcitedefaultseppunct}\relax
\EndOfBibitem
\bibitem[Perdew \latin{et~al.}(1996)Perdew, Burke, and Ernzerhof]{perdewGeneralizedGradientApproximation1996}
Perdew,~J.~P.; Burke,~K.; Ernzerhof,~M. Generalized {{Gradient Approximation Made Simple}}. \emph{Phys. Rev. Lett.} \textbf{1996}, \emph{77}, 3865--3868\relax
\mciteBstWouldAddEndPuncttrue
\mciteSetBstMidEndSepPunct{\mcitedefaultmidpunct}
{\mcitedefaultendpunct}{\mcitedefaultseppunct}\relax
\EndOfBibitem
\bibitem[Grimme \latin{et~al.}(2011)Grimme, Ehrlich, and Goerigk]{grimmeEffectDampingFunction2011}
Grimme,~S.; Ehrlich,~S.; Goerigk,~L. Effect of the Damping Function in Dispersion Corrected Density Functional Theory. \emph{J. Comput. Chem.} \textbf{2011}, \emph{32}, 1456--1465\relax
\mciteBstWouldAddEndPuncttrue
\mciteSetBstMidEndSepPunct{\mcitedefaultmidpunct}
{\mcitedefaultendpunct}{\mcitedefaultseppunct}\relax
\EndOfBibitem
\bibitem[Monkhorst and Pack(1976)Monkhorst, and Pack]{monkhorstSpecialPointsBrillouinzone1976}
Monkhorst,~H.~J.; Pack,~J.~D. Special Points for {{Brillouin-zone}} Integrations. \emph{Phys. Rev. B} \textbf{1976}, \emph{13}, 5188--5192\relax
\mciteBstWouldAddEndPuncttrue
\mciteSetBstMidEndSepPunct{\mcitedefaultmidpunct}
{\mcitedefaultendpunct}{\mcitedefaultseppunct}\relax
\EndOfBibitem
\bibitem[Manz and Limas(2016)Manz, and Limas]{manzIntroducingDDEC6Atomic2016}
Manz,~T.~A.; Limas,~N.~G. Introducing {{DDEC6}} Atomic Population Analysis: Part 1. {{Charge}} Partitioning Theory and Methodology. \emph{RSC Adv.} \textbf{2016}, \emph{6}, 47771--47801\relax
\mciteBstWouldAddEndPuncttrue
\mciteSetBstMidEndSepPunct{\mcitedefaultmidpunct}
{\mcitedefaultendpunct}{\mcitedefaultseppunct}\relax
\EndOfBibitem
\bibitem[Manz(2017)]{manzIntroducingDDEC6Atomic2017}
Manz,~T.~A. Introducing {{DDEC6}} Atomic Population Analysis: Part 3. {{Comprehensive}} Method to Compute Bond Orders. \emph{RSC Adv.} \textbf{2017}, \emph{7}, 45552--45581\relax
\mciteBstWouldAddEndPuncttrue
\mciteSetBstMidEndSepPunct{\mcitedefaultmidpunct}
{\mcitedefaultendpunct}{\mcitedefaultseppunct}\relax
\EndOfBibitem
\bibitem[Henkelman and J{\'o}nsson(2000)Henkelman, and J{\'o}nsson]{henkelmanImprovedTangentEstimate2000}
Henkelman,~G.; J{\'o}nsson,~H. Improved Tangent Estimate in the Nudged Elastic Band Method for Finding Minimum Energy Paths and Saddle Points. \emph{J. Chem. Phys.} \textbf{2000}, \emph{113}, 9978--9985\relax
\mciteBstWouldAddEndPuncttrue
\mciteSetBstMidEndSepPunct{\mcitedefaultmidpunct}
{\mcitedefaultendpunct}{\mcitedefaultseppunct}\relax
\EndOfBibitem
\bibitem[Henkelman \latin{et~al.}(2000)Henkelman, Uberuaga, and J{\'o}nsson]{henkelmanClimbingImageNudged2000}
Henkelman,~G.; Uberuaga,~B.~P.; J{\'o}nsson,~H. A Climbing Image Nudged Elastic Band Method for Finding Saddle Points and Minimum Energy Paths. \emph{J. Chem. Phys.} \textbf{2000}, \emph{113}, 9901--9904\relax
\mciteBstWouldAddEndPuncttrue
\mciteSetBstMidEndSepPunct{\mcitedefaultmidpunct}
{\mcitedefaultendpunct}{\mcitedefaultseppunct}\relax
\EndOfBibitem
\bibitem[Sun \latin{et~al.}(2015)Sun, Ruzsinszky, and Perdew]{sunStronglyConstrainedAppropriately2015}
Sun,~J.; Ruzsinszky,~A.; Perdew,~J.~P. Strongly {{Constrained}} and {{Appropriately Normed Semilocal Density Functional}}. \emph{Phys. Rev. Lett.} \textbf{2015}, \emph{115}, 036402\relax
\mciteBstWouldAddEndPuncttrue
\mciteSetBstMidEndSepPunct{\mcitedefaultmidpunct}
{\mcitedefaultendpunct}{\mcitedefaultseppunct}\relax
\EndOfBibitem
\bibitem[Fonseca~Guerra \latin{et~al.}(1998)Fonseca~Guerra, Snijders, {te Velde}, and Baerends]{fonsecaguerraOrderNDFTMethod1998}
Fonseca~Guerra,~C.; Snijders,~J.~G.; {te Velde},~G.; Baerends,~E.~J. Towards an Order-{{N DFT}} Method. \emph{Theor. Chem. Acc.} \textbf{1998}, \emph{99}, 391--403\relax
\mciteBstWouldAddEndPuncttrue
\mciteSetBstMidEndSepPunct{\mcitedefaultmidpunct}
{\mcitedefaultendpunct}{\mcitedefaultseppunct}\relax
\EndOfBibitem
\bibitem[{te Velde} \latin{et~al.}(2001){te Velde}, Bickelhaupt, Baerends, Fonseca~Guerra, {van Gisbergen}, Snijders, and Ziegler]{teveldeChemistryADF2001}
{te Velde},~G.; Bickelhaupt,~F.~M.; Baerends,~E.~J.; Fonseca~Guerra,~C.; {van Gisbergen},~S. J.~A.; Snijders,~J.~G.; Ziegler,~T. Chemistry with {{ADF}}. \emph{J. Comput. Chem.} \textbf{2001}, \emph{22}, 931--967\relax
\mciteBstWouldAddEndPuncttrue
\mciteSetBstMidEndSepPunct{\mcitedefaultmidpunct}
{\mcitedefaultendpunct}{\mcitedefaultseppunct}\relax
\EndOfBibitem
\bibitem[AMS()]{AMS2022}
{Rüger, R.; Franchini, M.; Trnka, T.; Yakovlev, A.; Philipsen, P.; van Vuren, T., Klumpers, B., Soini, T.}, \textit{AMS 2022}, {SCM, Theoretical Chemistry, Vrije Universiteit, Amsterdam, The Netherlands}, {2022}\relax
\mciteBstWouldAddEndPuncttrue
\mciteSetBstMidEndSepPunct{\mcitedefaultmidpunct}
{\mcitedefaultendpunct}{\mcitedefaultseppunct}\relax
\EndOfBibitem
\bibitem[van Lenthe and Baerends(2003)van Lenthe, and Baerends]{lentheOptimizedSlatertypeBasis2003}
van Lenthe,~E.; Baerends,~E.~J. Optimized {{Slater-type}} Basis Sets for the Elements 1\textendash 118. \emph{J. Comput. Chem.} \textbf{2003}, \emph{24}, 1142--1156\relax
\mciteBstWouldAddEndPuncttrue
\mciteSetBstMidEndSepPunct{\mcitedefaultmidpunct}
{\mcitedefaultendpunct}{\mcitedefaultseppunct}\relax
\EndOfBibitem
\bibitem[van Lenthe \latin{et~al.}(1993)van Lenthe, Baerends, and Snijders]{lentheRelativisticRegularTwo1993}
van Lenthe,~E.; Baerends,~E.~J.; Snijders,~J.~G. Relativistic Regular Two-component {{Hamiltonians}}. \emph{J. Chem. Phys.} \textbf{1993}, \emph{99}, 4597--4610\relax
\mciteBstWouldAddEndPuncttrue
\mciteSetBstMidEndSepPunct{\mcitedefaultmidpunct}
{\mcitedefaultendpunct}{\mcitedefaultseppunct}\relax
\EndOfBibitem
\bibitem[Hansen and Ostermeier(2001)Hansen, and Ostermeier]{hansenCompletelyDerandomizedSelfAdaptation2001}
Hansen,~N.; Ostermeier,~A. Completely {{Derandomized Self-Adaptation}} in {{Evolution Strategies}}. \emph{Evol. Comput.} \textbf{2001}, \emph{9}, 159--195\relax
\mciteBstWouldAddEndPuncttrue
\mciteSetBstMidEndSepPunct{\mcitedefaultmidpunct}
{\mcitedefaultendpunct}{\mcitedefaultseppunct}\relax
\EndOfBibitem
\bibitem[Mortier \latin{et~al.}(1986)Mortier, Ghosh, and Shankar]{mortierElectronegativityequalizationMethodCalculation1986}
Mortier,~W.~J.; Ghosh,~S.~K.; Shankar,~S. Electronegativity-Equalization Method for the Calculation of Atomic Charges in Molecules. \emph{J. Am. Chem. Soc.} \textbf{1986}, \emph{108}, 4315--4320\relax
\mciteBstWouldAddEndPuncttrue
\mciteSetBstMidEndSepPunct{\mcitedefaultmidpunct}
{\mcitedefaultendpunct}{\mcitedefaultseppunct}\relax
\EndOfBibitem
\bibitem[Chen \latin{et~al.}(2021)Chen, Brocks, Tao, and Bobbert]{chenUnifiedTheoryLightinduced2021}
Chen,~Z.; Brocks,~G.; Tao,~S.; Bobbert,~P.~A. Unified Theory for Light-Induced Halide Segregation in Mixed Halide Perovskites. \emph{Nat. Commun.} \textbf{2021}, \emph{12}, 2687\relax
\mciteBstWouldAddEndPuncttrue
\mciteSetBstMidEndSepPunct{\mcitedefaultmidpunct}
{\mcitedefaultendpunct}{\mcitedefaultseppunct}\relax
\EndOfBibitem
\bibitem[Chen \latin{et~al.}(2022)Chen, Guo, Gong, and Yin]{chenKineticPathwayGtod2022}
Chen,~G.-Y.; Guo,~Z.-D.; Gong,~X.-G.; Yin,~W.-J. Kinetic Pathway of {\textgamma}-to-{\textdelta} Phase Transition in {\ch{CsPbI3}}. \emph{Chem} \textbf{2022}, \emph{8}, 3120--3129\relax
\mciteBstWouldAddEndPuncttrue
\mciteSetBstMidEndSepPunct{\mcitedefaultmidpunct}
{\mcitedefaultendpunct}{\mcitedefaultseppunct}\relax
\EndOfBibitem
\bibitem[Schl{\"u}ter and Schl{\"u}ter(1974)Schl{\"u}ter, and Schl{\"u}ter]{schluterElectronicStructureOptical1974}
Schl{\"u}ter,~I.~{\relax Ch}.; Schl{\"u}ter,~M. Electronic Structure and Optical Properties of \ch{PbI2}. \emph{Phys. Rev. B} \textbf{1974}, \emph{9}, 1652--1663\relax
\mciteBstWouldAddEndPuncttrue
\mciteSetBstMidEndSepPunct{\mcitedefaultmidpunct}
{\mcitedefaultendpunct}{\mcitedefaultseppunct}\relax
\EndOfBibitem
\bibitem[Karmakar \latin{et~al.}(2019)Karmakar, Dodd, Zhang, Oakley, Klobukowski, and Michaelis]{karmakarMechanochemicalSynthesis0D2019}
Karmakar,~A.; Dodd,~M.~S.; Zhang,~X.; Oakley,~M.~S.; Klobukowski,~M.; Michaelis,~V.~K. Mechanochemical Synthesis of {{0D}} and {{3D}} Cesium Lead Mixed Halide Perovskites. \emph{Chem. Commun.} \textbf{2019}, \emph{55}, 5079--5082\relax
\mciteBstWouldAddEndPuncttrue
\mciteSetBstMidEndSepPunct{\mcitedefaultmidpunct}
{\mcitedefaultendpunct}{\mcitedefaultseppunct}\relax
\EndOfBibitem
\bibitem[Berendsen \latin{et~al.}(1984)Berendsen, Postma, {van Gunsteren}, DiNola, and Haak]{berendsenMolecularDynamicsCoupling1984}
Berendsen,~H. J.~C.; Postma,~J. P.~M.; {van Gunsteren},~W.~F.; DiNola,~A.; Haak,~J.~R. Molecular Dynamics with Coupling to an External Bath. \emph{J. Chem. Phys.} \textbf{1984}, \emph{81}, 3684--3690\relax
\mciteBstWouldAddEndPuncttrue
\mciteSetBstMidEndSepPunct{\mcitedefaultmidpunct}
{\mcitedefaultendpunct}{\mcitedefaultseppunct}\relax
\EndOfBibitem
\bibitem[Martyna \latin{et~al.}(1992)Martyna, Klein, and Tuckerman]{martynaNoseHooverChains1992}
Martyna,~G.~J.; Klein,~M.~L.; Tuckerman,~M. Nos\'e\textendash{{Hoover}} Chains: {{The}} Canonical Ensemble via Continuous Dynamics. \emph{J. Chem. Phys.} \textbf{1992}, \emph{97}, 2635--2643\relax
\mciteBstWouldAddEndPuncttrue
\mciteSetBstMidEndSepPunct{\mcitedefaultmidpunct}
{\mcitedefaultendpunct}{\mcitedefaultseppunct}\relax
\EndOfBibitem
\bibitem[Martyna \latin{et~al.}(1994)Martyna, Tobias, and Klein]{martynaConstantPressureMolecular1994}
Martyna,~G.~J.; Tobias,~D.~J.; Klein,~M.~L. Constant Pressure Molecular Dynamics Algorithms. \emph{J. Chem. Phys.} \textbf{1994}, \emph{101}, 4177--4189\relax
\mciteBstWouldAddEndPuncttrue
\mciteSetBstMidEndSepPunct{\mcitedefaultmidpunct}
{\mcitedefaultendpunct}{\mcitedefaultseppunct}\relax
\EndOfBibitem
\bibitem[Larsen \latin{et~al.}(2016)Larsen, Schmidt, and Schi{\o}tz]{larsenRobustStructuralIdentification2016}
Larsen,~P.~M.; Schmidt,~S.; Schi{\o}tz,~J. Robust Structural Identification via Polyhedral Template Matching. \emph{Model. Simul. Mater. Sci. Eng.} \textbf{2016}, \emph{24}, 055007\relax
\mciteBstWouldAddEndPuncttrue
\mciteSetBstMidEndSepPunct{\mcitedefaultmidpunct}
{\mcitedefaultendpunct}{\mcitedefaultseppunct}\relax
\EndOfBibitem
\bibitem[Stukowski(2009)]{stukowskiVisualizationAnalysisAtomistic2009}
Stukowski,~A. Visualization and Analysis of Atomistic Simulation Data with {OVITO}-{The} Open Visualization Tool. \emph{Model. Simul. Mater. Sci. Eng.} \textbf{2009}, \emph{18}, 015012\relax
\mciteBstWouldAddEndPuncttrue
\mciteSetBstMidEndSepPunct{\mcitedefaultmidpunct}
{\mcitedefaultendpunct}{\mcitedefaultseppunct}\relax
\EndOfBibitem
\bibitem[Wiktor \latin{et~al.}(2023)Wiktor, Fransson, Kubicki, and Erhart]{wiktorQuantifyingDynamicTilting2023}
Wiktor,~J.; Fransson,~E.; Kubicki,~D.; Erhart,~P. Quantifying {{Dynamic Tilting}} in {{Halide Perovskites}}: {{Chemical Trends}} and {{Local Correlations}}. \emph{Chem. Mater.} \textbf{2023}, \emph{35}, 6737--6744\relax
\mciteBstWouldAddEndPuncttrue
\mciteSetBstMidEndSepPunct{\mcitedefaultmidpunct}
{\mcitedefaultendpunct}{\mcitedefaultseppunct}\relax
\EndOfBibitem
\bibitem[Goldschmidt(1926)]{goldschmidtGesetzeKrystallochemie1926}
Goldschmidt,~V.~M. {Die Gesetze der Krystallochemie}. \emph{Naturwissenschaften} \textbf{1926}, \emph{14}, 477--485\relax
\mciteBstWouldAddEndPuncttrue
\mciteSetBstMidEndSepPunct{\mcitedefaultmidpunct}
{\mcitedefaultendpunct}{\mcitedefaultseppunct}\relax
\EndOfBibitem
\bibitem[Shannon(1976)]{shannonRevisedEffectiveIonic1976}
Shannon,~R.~D. Revised Effective Ionic Radii and Systematic Studies of Interatomic Distances in Halides and Chalcogenides. \emph{Acta Crystallogr. A.} \textbf{1976}, \emph{32}, 751--767\relax
\mciteBstWouldAddEndPuncttrue
\mciteSetBstMidEndSepPunct{\mcitedefaultmidpunct}
{\mcitedefaultendpunct}{\mcitedefaultseppunct}\relax
\EndOfBibitem
\end{mcitethebibliography}

\end{document}


\clearpage

\tableofcontents

\clearpage



\section*{Supporting Notes}

\subsection{Reference data}

The reference data used in the training set was generated using density functional theory (DFT) calculations. Both periodic and molecular structures were included in the training set, for each type of structures the details are provided in the following section.

\subsubsection{Periodic systems}

Calculations on periodic structures were done with the projector-augmented wave (PAW) method~\cite{blochlProjectorAugmentedwaveMethod1994, kresseUltrasoftPseudopotentialsProjector1999} as implemented in the Vienna Ab-initio Simulation Package (VASP)~\cite{kresseInitioMoleculardynamicsSimulation1994, kresseEfficiencyAbinitioTotal1996, kresseEfficientIterativeSchemes1996}. In line with previous work~\cite{polsAtomisticInsightsDegradation2021}, most of the reference data was calculated using the PBE exchange-correlation (XC) functional~\cite{perdewGeneralizedGradientApproximation1996} with long-range dispersion interactions accounted for by the DFT-D3(BJ) dispersion correction~\cite{grimmeEffectDampingFunction2011}. The outermost electrons of \ch{Br} (4s$^{2}$4p$^{5}$), \ch{I} (5s$^{2}$5p$^{5}$), \ch{Cs} (5s$^{2}$5p$^{6}$6s$^{1}$) and \ch{Pb} (5d$^{10}$6s$^{2}$6p$^{2}$) were treated as valence electrons. In the calculations we used a plane-wave cutoff energy of \SI{500}{\eV} together with regularly spaced meshes for the Brillouin zone integration~\cite{monkhorstSpecialPointsBrillouinzone1976}.  An overview of the $k$-meshes used for these calculations can be found in Table~\ref{tab:kpoints_eos}.

The ground state structures of the precursors and perovskites were obtained by allowing all the ionic positions, cell shape and cell volume to change. During the geometry optimizations, energy and force convergence criteria of \SI{1E-2}{\meV} and \SI{10}{\meV\per\angstrom}, respectively, were used. The atomic charges in these materials were calculated with the DDEC6 method~\cite{manzIntroducingDDEC6Atomic2016, manzIntroducingDDEC6Atomic2017}. Whenever monolayers or slabs were modeled, vacuum layers of at least \SI{15}{\angstrom} were inserted between the atoms to prevent interactions between the periodic images. Equations of state were generated by applying a strain to the equilibrium geometries in the direction of the lattice vectors and subsequently allowing for the ionic positions to relax using the above-mentioned convergence criteria.

Defect formation energies were calculated for monolayers of \ch{PbI2} and orthorhombic \ch{CsPbX3} (\ch{X} = \ch{Br}/\ch{I}) by taking the difference in energy between the defective and pristine structures. To prevent interactions between the periodic copies of the defects, we used \supercell{4}{4}{1} and \supercell{2}{2}{1} supercells for \ch{PbI2} monolayers and orthorhombic \ch{CsPbX3}, respectively. Accordingly, the $k$-meshes used to sample reciprocal space were scaled to \kmesh{3}{3}{1} and \kmesh{2}{2}{3}. Lower convergence criteria were used to lower the cost of the defect calculations, \SI{1E-1}{\meV} and \SI{30}{\meV\per\angstrom} for \ch{PbI2} and \SI{1E-2}{\meV} and \SI{50}{\meV\per\angstrom} for \ch{CsPbX3}.

Defect migration barriers were calculated using transition state calculations with the Climbing Image Nudged Elastic Band (CI-NEB) method~\cite{henkelmanImprovedTangentEstimate2000, henkelmanClimbingImageNudged2000}. For all barriers five intermediate geometries were used for the calculations. The migration barriers in \ch{PbI2} monolayers were calculated in a \supercell{4}{4}{1} supercell with a \kmesh{3}{3}{1} $k$-mesh, for computational efficiency the vacuum was reduced to \SI{10}{\angstrom} here. For \ch{CsPbX3} perovskites, we used a \supercell{3}{3}{3} cubic supercell sampled with only the $\Gamma$-point (\kmesh{1}{1}{1} $k$-mesh). The defect migration barriers in perovskites were calculated using the SCAN XC functional~\cite{sunStronglyConstrainedAppropriately2015} and made use of a \ch{Pb} pseudopotential with fewer valence electrons (6s$^{2}$6p$^{2}$).

\begin{table}[ht]
    \caption{The $k$-meshes used for the different materials for which an equations of state was calculated. \ch{X} is used to denote halides (\ch{Br} or \ch{I}).}
    \label{tab:kpoints_eos}
    \begin{tabular}{@{}ccc@{}}
        \toprule
        Material                                & Phase                     & $k$-points            \\ \midrule
        \multirow{4}{*}{\ch{CsPbX3}}            & Cubic (\textalpha)        & \kmesh{6}{6}{6}       \\
                                                & Tetragonal (\textbeta)    & \kmesh{4}{4}{6}       \\
                                                & Orthorhombic (\textgamma) & \kmesh{4}{4}{3}       \\
                                                & Yellow (\textdelta)       & \kmesh{6}{3}{2}       \\ \midrule
        \multirow{2}{*}{\ch{CsX}}               & \ch{CsCl}-type            & \kmesh{12}{12}{12}    \\
                                                & \ch{NaCl}-type            & \kmesh{8}{8}{8}       \\ \midrule
        \multirow{2}{*}{\ch{PbI2}}              & Hexagonal                 & \kmesh{11}{11}{7}     \\
                                                & \ch{MoS2}-type            & \kmesh{11}{11}{4}     \\ \midrule
        \multirow{3}{*}{\ch{PbI2} monolayer}    & Hexagonal monolayer       & \kmesh{11}{11}{1}     \\
                                                & \ch{MoS2}-type monolayer  & \kmesh{11}{11}{1}     \\
                                                & Octagonal monolayer       & \kmesh{12}{12}{1}     \\ \midrule
        \ch{PbBr2}                              & \ch{PbCl2}-type           & \kmesh{5}{11}{5}      \\ \bottomrule
    \end{tabular}
\end{table}

\subsubsection{Molecular systems}

Molecular geometries \ch{PbX2} (\ch{X} = \ch{Br}/\ch{I}) were evaluated using DFT calculations with ADF~\cite{fonsecaguerraOrderNDFTMethod1998, teveldeChemistryADF2001} in AMS2022~\cite{AMS2022}. For all species (\ch{Br}, \ch{I} and \ch{Pb}) all electrons were treated as valence electrons. The basis set for the calculations was a TZP basis set of Slater Type Orbitals (STO)~\cite{lentheOptimizedSlatertypeBasis2003}. Scalar relativistic effects were included using the Zero Order Regular Approximated (ZORA) Hamiltonian~\cite{lentheRelativisticRegularTwo1993}. Analogous to the majority of the bulk calculations, the PBE XC functional~\cite{perdewGeneralizedGradientApproximation1996} was used to model the electronic exchange-correlation interactions. Dispersion interactions were accounted for by the Becke-Johnson damped DFT-D3 scheme~\cite{grimmeEffectDampingFunction2011}. Consequently, these settings were used to scan the potential energy surface of \ch{PbX2} molecular systems. Using constrained geometry optimizations the energy profile of the $\ch{X}-\ch{Pb}-\ch{X}$ valence angles and $\ch{Pb}-\ch{X}$ bonds were scanned.

\clearpage

\subsection{Force field optimization}

\subsubsection{Parameter optimization}

The parameter optimization was carried out using a covariance matrix adaptation evolutionary strategy (CMA-ES) algorithm proposed by \citet{hansenCompletelyDerandomizedSelfAdaptation2001}. During parameter optimization a set of pre-defined parameters in the ReaxFF force field was allowed to change within set bounds. During each optimization step, a collection of new parameter sets was generated. The quality of the parameter sets in this collection was evaluated using the sum of squared errors (SSE) loss function constructed from the entries in the training set. During successive generations, the distribution of parameter sets is improved upon by moving into the direction of the lowest values for the loss function. We made use of a population size of $N_{\mathrm{pop}} = 16$ for each generation, moreover we used an initial standard deviation for the multivariate normal distribution of $\sigma = 0.2$ from which trial parameter sets could be drawn.

\subsubsection{Parameter scaling}

The \ch{Br} parameters were initialized from scaled atomic and interatomic \ch{I} parameters as found in the retrained \oldspecies{} ReaxFF force field. The scaling constants used for the procedure are reported in Table~\ref{tab:si_parameter_scaling}. The \ch{Br} parameters were allowed to change during the parameter optimization procedure. To adhere to the chemical trends for ion size and bond distances, we applied constraints during the parameter optimization. With the constraints, shown in Table~\ref{tab:si_parameter_constraints}, we ensured that Van der Waals interactions are longer-ranged than covalent interactions ($r_{\mathrm{vdW}}$ $>$ $r_{0}^{\mathrm{sigma}}$) and that all \ch{I}-containing interactions had a longer range than their equivalent \ch{Br}-containing interactions.

\begin{table}[htbp!]
    \caption{Scale constants used to convert \ch{I} atomic and interatomic parameters to parameters for \ch{Br}.}
    \label{tab:si_parameter_scaling}
    \begin{tabular}{@{}cccc@{}}
        \toprule
        Interaction                         & Parameter                                             & Elements          & Scale constants   \\ \midrule
        \multirow{4}{*}{Bonding (BND)}      & \multirow{4}{*}{$D_{\mathrm{e}}^{\mathrm{sigma}}$}    & \ch{Br}.\ch{I}    & 1.05              \\
                                            &                                                       & \ch{Br}.\ch{Br}   & 1.10              \\
                                            &                                                       & \ch{Br}.\ch{Pb}   & 1.10              \\
                                            &                                                       & \ch{Br}.\ch{Cs}   & 1.10              \\ \midrule
        \multirow{6}{*}{Off-diagonal (OFD)} & \multirow{2}{*}{$D_{\mathrm{ij}}$}                    & \ch{Br}.\ch{Pb}   & 1.10              \\
                                            &                                                       & \ch{Br}.\ch{Cs}   & 1.10              \\ \cmidrule(l){2-4} 
                                            & \multirow{2}{*}{$r_{\mathrm{vdW}}$}                   & \ch{Br}.\ch{Pb}   & 0.95              \\
                                            &                                                       & \ch{Br}.\ch{Cs}   & 0.95              \\ \cmidrule(l){2-4} 
                                            & \multirow{2}{*}{$r_{0}^{\mathrm{sigma}}$}             & \ch{Br}.\ch{Pb}   & 0.95              \\
                                            &                                                       & \ch{Br}.\ch{Cs}   & 0.95              \\ \bottomrule
    \end{tabular}
\end{table}

\begin{table}[htbp!]
    \caption{Scale constants used to convert \ch{I} atomic and interatomic parameters to parameters for \ch{Br}.}
    \label{tab:si_parameter_constraints}
    \begin{tabular}{@{}cc@{}}
        \toprule
                            & Constraint                                                                            \\ \midrule
        1.                  & \ch{Cs}.\ch{I}:$r_{\mathrm{vdW}}$ $>$ \ch{Cs}.\ch{I}:$r_{0}^{\mathrm{sigma}}$         \\
        2.                  & \ch{Br}.\ch{Cs}:$r_{\mathrm{vdW}}$ $>$ \ch{Br}.\ch{Cs}:$r_{0}^{\mathrm{sigma}}$       \\
        3.                  & \ch{I}.\ch{Pb}:$r_{\mathrm{vdW}}$ $>$ \ch{I}.\ch{Pb}:$r_{0}^{\mathrm{sigma}}$         \\
        4.                  & \ch{Br}.\ch{Pb}:$r_{\mathrm{vdW}}$ $>$ \ch{Br}.\ch{Pb}:$r_{0}^{\mathrm{sigma}}$       \\
        5.                  & \ch{Cs}.\ch{I}:$r_{\mathrm{vdW}}$ $>$ \ch{Br}.\ch{Cs}:$r_{\mathrm{vdW}}$              \\
        6.                  & \ch{Cs}.\ch{I}:$r_{0}^{\mathrm{sigma}}$ $>$ \ch{Br}.\ch{Cs}:$r_{0}^{\mathrm{sigma}}$  \\
        7.                  & \ch{I}.\ch{Pb}:$r_{\mathrm{vdW}}$ $>$ \ch{Br}.\ch{Pb}:$r_{\mathrm{vdW}}$              \\
        8.                  & \ch{I}.\ch{Pb}:$r_{0}^{\mathrm{sigma}}$ $>$ \ch{Br}.\ch{Pb}:$r_{0}^{\mathrm{sigma}}$  \\ \bottomrule
    \end{tabular}
\end{table}

\clearpage

\subsection{Force field validation}

\subsubsection{Atomic charges}

The electronegativity equalization method (EEM)~\cite{mortierElectronegativityequalizationMethodCalculation1986} is used to determine the atomic charges in the ReaxFF force field. For DFT calculations the charges were obtained using the DDEC6 method~\cite{manzIntroducingDDEC6Atomic2016, manzIntroducingDDEC6Atomic2017}. We note that the new ReaxFF parameters accurately reproduce the atomic charges of different species in a variety of compounds, which include precursor and perovskite geometries. A comparison of the charges from DFT and from ReaxFF is provided in Table~\ref{tab:si_atomic_charges}.

\begin{table}[htbp!]
    \caption{Atomic charges of different species in a variety of compounds as compared between DFT and ReaxFF calculations.}
    \label{tab:si_atomic_charges}
    \begin{tabular}{@{}cccccc@{}}
        \toprule
        \multirow{2}{*}{Compound}               & \multirow{2}{*}{Element}  & \multicolumn{2}{c}{\ch{X} = \ch{I}}                           & \multicolumn{2}{c}{\ch{X} = \ch{Br}}                          \\
                                                &                           & $q_{\mathrm{DFT}}$ ($e$)      & $q_{\mathrm{ReaxFF}}$ ($e$)   & $q_{\mathrm{DFT}}$ ($e$)      & $q_{\mathrm{ReaxFF}}$ ($e$)   \\ \midrule
        \multirow{2}{*}{\ch{CsX}}               & \ch{Cs}                   & $+0.80$                       & $+0.72$                       & $+0.82$                       & $+0.77$                       \\
                                                & \ch{X}                    & $-0.80$                       & $-0.72$                       & $-0.82$                       & $-0.77$                       \\ \midrule
        \multirow{2}{*}{\ch{PbX2}}              & \ch{Pb}                   & $+0.55$                       & $+0.71$                       & $+0.90$                       & $+0.88$                       \\
                                                & \ch{X}                    & $-0.28$                       & $-0.36$                       & $-0.45$                       & $-0.44$                       \\ \midrule
        \multirow{3}{*}{\textalpha-\ch{CsPbX3}} & \ch{Cs}                   & $+0.83$                       & $+0.80$                       & $+0.84$                       & $+0.85$                       \\
                                                & \ch{Pb}                   & $+0.81$                       & $+0.69$                       & $+0.95$                       & $+0.73$                       \\
                                                & \ch{X}                    & $-0.55$                       & $-0.49$                       & $-0.60$                       & $-0.53$                       \\ \midrule
        \multirow{3}{*}{\textbeta-\ch{CsPbX3}}  & \ch{Cs}                   & $+0.81$                       & $+0.81$                       & $+0.83$                       & $+0.86$                       \\
                                                & \ch{Pb}                   & $+0.80$                       & $+0.67$                       & $+0.94$                       & $+0.73$                       \\
                                                & \ch{X}                    & $-0.53$                       & $-0.49$                       & $-0.58$                       & $-0.53$                       \\ \midrule
        \multirow{4}{*}{\textgamma-\ch{CsPbX3}} & \ch{Cs}                   & $+0.78$                       & $+0.83$                       & $+0.81$                       & $+0.87$                       \\
                                                & \ch{Pb}                   & $+0.78$                       & $+0.67$                       & $+0.94$                       & $+0.74$                       \\
                                                & axial \ch{X}              & $-0.52$                       & $-0.49$                       & $-0.58$                       & $-0.54$                       \\
                                                & equatorial \ch{X}         & $-0.52$                       & $-0.50$                       & $-0.58$                       & $-0.53$                       \\ \bottomrule
    \end{tabular}
\end{table}

\clearpage

\subsubsection{Equations of state}

\begin{figure*}[htbp!]
    \includegraphics{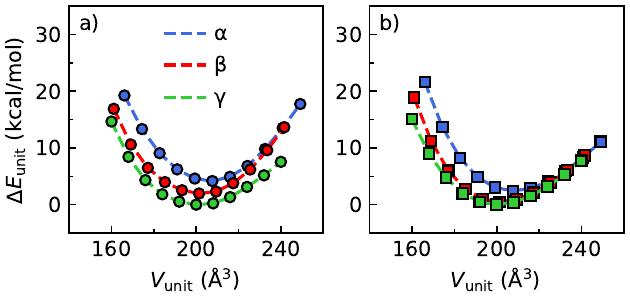}
    \caption{Equations of state of various phases of \ch{CsPbBr3} from a) ReaxFF and b) DFT calculations.}
    \label{fig:si_equations_of_state}
\end{figure*}

In addition to the equations of state for \ch{CsPbI3} shown in the main text, we show that the ReaxFF force field also captures the bulk phase behavior of \ch{CsPbBr3} in Figure~\ref{fig:si_equations_of_state}. We note that the ReaxFF parameter appropriately ranks the various perovskite phases of \ch{CsPbBr3} from least to most stable as: \textalpha~$<$ \textbeta~$<$ \textdelta. The relative energies of the different perovskite and nonperovskite phases are provided in Table~\ref{tab:si_phase_energies}.

\begin{table}[htbp!]
    \caption{Relative energies in \SI{}{\kcal\per\mol} of the various phases of \ch{CsPbI3} and \ch{CsPbBr3} with respect to the \textgamma-phase.}
    \label{tab:si_phase_energies}
    \begin{tabular}{@{}ccccc@{}}
        \toprule
        \multirow{2}{*}{Phase}  & \multicolumn{2}{c}{\ch{CsPbI3}}       & \multicolumn{2}{c}{\ch{CsPbBr3}}      \\
                                & DFT               & ReaxFF            & DFT               & ReaxFF            \\ \midrule
        \textalpha              & $+3.54$           & $+2.88$           & $+2.43$           & $+4.16$           \\
        \textbeta               & $+0.96$           & $+1.33$           & $+0.49$           & $+1.96$           \\
        \textgamma              & $+0.00$           & $+0.00$           & $+0.00$           & $+0.00$           \\
        \textdelta              & $-3.06$           & $-5.87$           & -                 & -                 \\ \bottomrule
    \end{tabular}
\end{table}

\clearpage

\subsubsection{Mixing enthalpies}

The mixing enthalpy of \ch{CsPb(Br_{x}I_{1-x})_{3}} perovskites was determined by comparing the total energy of mixed halide perovskites to that of the pure perovskites (i.e. \ch{CsPbI3} and \ch{CsPbBr3}). The following equation was used to calculate the mixing enthalpies
\begin{equation}
    \label{eq:mixing_enthalpies}
    E^{\mathrm{mix}}_{\mathrm{unit}} = E^{\ch{CsPb(Br_{x}I_{1-x})_{3}}}_{\mathrm{unit}} - x \cdot E^{\ch{CsPbBr3}}_{\mathrm{unit}} - (1 - x) \cdot E^{\ch{CsPbI3}}_{\mathrm{unit}}
\end{equation}
here $E_{\mathrm{unit}}$ is the total energies per formula unit for \ch{CsPb(Br_{x}I_{1-x})_{3}}, \ch{CsPbBr3} and \ch{CsPbI3}. For each mixed halide perovskite, the lowest energy structure from a set of six randomly mixed orthorhombic structures was used to determine the mixing enthalpy. Due to this sampling, it is possible that an outlier geometry was selected in the determination of the mixing enthalpy (e.g. only axial substitutions for x = 1/6 and x = 1/4), thus resulting in a too low mixing enthalpy. The resulting mixing enthalpies, shown in Table~\ref{tab:si_mixing_enthalpies}, corroborate previous DFT calculations in literature in which mixing enthalpies below \SI{1.0}{\kcal\per\mol} per formula unit were also found~\cite{chenUnifiedTheoryLightinduced2021}.

\begin{table}[htbp!]
    \caption{Mixing enthalpies of \ch{CsPb(Br_{x}I_{1-x})_{3}} perovskites determined with ReaxFF and DFT calculations in \SI{}{\kcal\per\mol}.}
    \label{tab:si_mixing_enthalpies}
    \begin{tabular}{@{}cccc@{}}
        \toprule 
        x         & x           & DFT                 & ReaxFF            \\ \midrule
        0         & 0/12        & $+0.00$             & $+0.00$           \\ 
        1/12      & 1/12        & $+0.14$             & $+0.15$           \\ 
        1/6       & 2/12        & $+0.08$             & $-0.39$           \\ 
        1/4       & 3/12        & $+0.11$             & $-0.10$           \\ 
        1/3       & 4/12        & $+0.50$             & $+0.53$           \\ 
        5/12      & 5/12        & $+0.71$             & $+0.83$           \\ 
        1/2       & 6/12        & $+0.68$             & $+0.64$           \\ 
        7/12      & 7/12        & $+0.55$             & $+0.51$           \\ 
        2/3       & 8/12        & $+0.39$             & $+0.56$           \\ 
        3/4       & 9/12        & $+0.47$             & $+0.37$           \\ 
        5/6       & 10/12       & $+0.26$             & $-0.04$           \\ 
        11/12     & 11/12       & $+0.19$             & $-0.03$           \\ 
        1         & 12/12       & $+0.00$             & $+0.00$           \\ \bottomrule
    \end{tabular}
\end{table}

\clearpage

\subsubsection{\ch{CsPbI3} degradation pathway}

\begin{figure*}[htbp!]
    \includegraphics{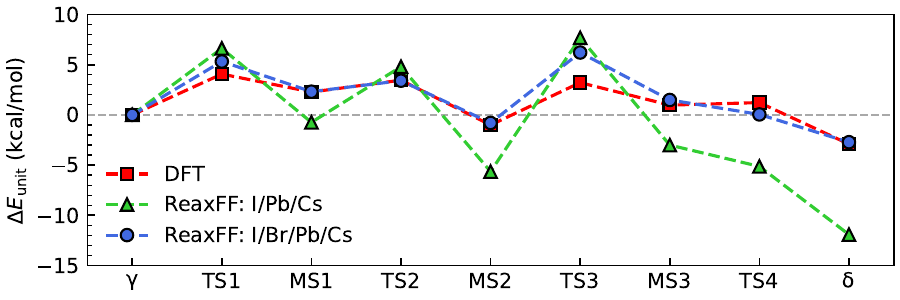}
    \caption{\ch{CsPbI3} degradation mechanism from the \textgamma-phase to the \textdelta-phase. Reference data from DFT calculations (squares) are compared to the \oldspecies{} (triangles) and \newspecies{} (circles) ReaxFF parameters.}
    \label{fig:si_gamma_delta_transition}
\end{figure*}

To check if the new ReaxFF force field improves on the description of the commonly observed degradation of \ch{CsPbI3}, we compared the \oldspecies{} and \newspecies{} ReaxFF parameters against the degradation mechanism of \ch{CsPbI3} from the \textgamma-phase to the \textdelta-phase as proposed by \citet{chenKineticPathwayGtod2022}. This comparison is shown in Figure~\ref{fig:si_gamma_delta_transition}. Notably, we find that the new set of ReaxFF parameters shows important improvements for the various metastable states (MS1, MS2 and MS3) and final state (\textdelta) along the reaction coordinate, which are overstabilized anywhere from \SI{3}{\kcal\per\mol} to \SI{9}{\kcal\per\mol} per formula unit with the \oldspecies{} parameter set. Moreover, we find that the reparameterized ReaxFF parameters provide more reasonable energies for some transition states (TS1 and TS3), the values of which are closer to the reference values from DFT calculations by about \SI{1.0}{\kcal\per\mol} to \SI{1.5}{\kcal\per\mol} per formula unit. A complete overview of all relative energies is shown in Table~\ref{tab:si_gamma_delta_transition}.

\begin{table}[htbp!]
    \caption{Relative energies per unit cell for DFT, \oldspecies{} ReaxFF and \newspecies{} ReaxFF calculations. Energies are provided in \SI{}{\kcal\per\mol}.}
    \label{tab:si_gamma_delta_transition}
    \begin{tabular}{@{}cccc@{}}
        \toprule
        \multirow{2}{*}{Phase}  & \multirow{2}{*}{DFT}      & ReaxFF:                       & ReaxFF:                   \\ 
                                &                           & \oldspecies{}                 & \newspecies{}             \\ \midrule
        \textgamma              & $+0.00$                   & $+0.00$                       & $+0.00$                   \\
        TS1                     & $+4.07$                   & $+6.62$                       & $+5.32$                   \\
        MS1                     & $+2.28$                   & $-0.74$                       & $+2.32$                   \\
        TS2                     & $+3.46$                   & $+4.80$                       & $+3.41$                   \\
        MS2                     & $-1.08$                   & $-5.81$                       & $-0.94$                   \\
        TS3                     & $+3.22$                   & $+7.69$                       & $+6.21$                   \\
        MS3                     & $+0.97$                   & $-3.01$                       & $+1.49$                   \\
        TS4                     & $+1.23$                   & $-5.14$                       & $+0.05$                   \\
        \textdelta              & $-2.91$                   & $-11.92$                      & $-2.72$                   \\ \bottomrule
    \end{tabular}
\end{table}

\clearpage

\subsubsection{Defect migration barriers}

\begin{figure*}[htbp!]
    \includegraphics{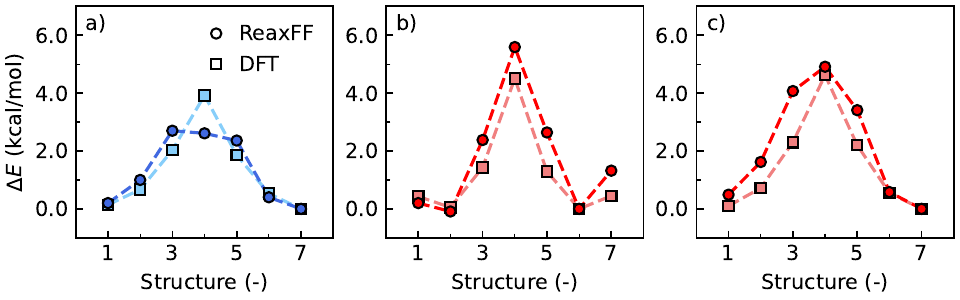}
    \caption{Defect migration barriers of a) \ch{I} interstitial in \ch{CsPbI3}, b) \ch{Br} vacancy in \ch{CsPbBr3} and c) \ch{Br} interstitial in \ch{CsPbBr3}. Data from the ReaxFF force field are shown in circles and from DFT calculations in squares.}
    \label{fig:si_defect_migration_barriers}
\end{figure*}

In addition to \ch{I} vacancies in \ch{CsPbI3}, we also calculated the migration barrier for other types of defects in inorganic halide perovskites as shown in Figure~\ref{fig:si_defect_migration_barriers}; \ch{I} interstitial in \ch{CsPbI3}, \ch{Br} vacancy in \ch{CsPbBr3} and \ch{Br} interstitial in \ch{CsPbBr3}. We note that all defect migration barriers match the DFT migration barriers to within \SI{1.5}{\kcal\per\mol}.

\clearpage

\subsubsection{\ch{PbX2} molecular geometries}

\begin{figure*}[htbp!]
    \includegraphics{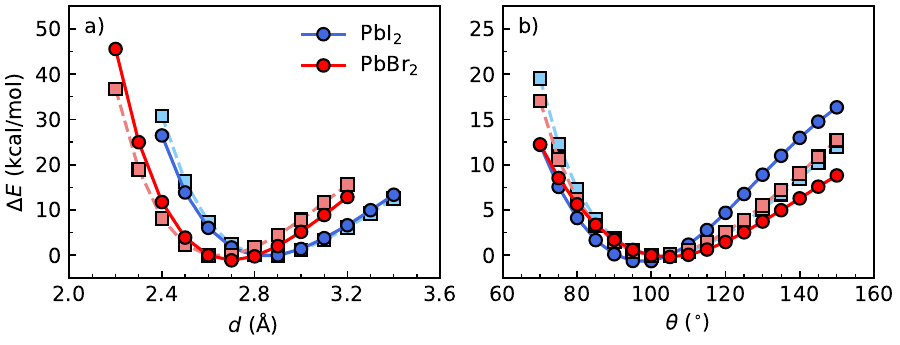}
    \caption{Geometry scans of \ch{PbI2} and \ch{PbBr2} molecules. a) $\ch{Pb}-\ch{X}$ bond distance scan and b) $\ch{X}-\ch{Pb}-\ch{X}$ valence angle scan. Data from the ReaxFF force field are shown in circles and from DFT calculations in squares.}
    \label{fig:si_pbx_geometries}
\end{figure*}

To validate that the ReaxFF parameter set works not only for bulk structures, but also molecular fragments, we benchmarked the parameter set against some \ch{PbX2} molecular geometries. For both \ch{PbI2} and \ch{PbBr2} we scanned the $\ch{Pb}-\ch{X}$ bond distance and $\ch{X}-\ch{Pb}-\ch{X}$ valence angle. The results of these scans can be found in Figure~\ref{fig:si_pbx_geometries} and show a good agreement between ReaxFF and the DFT reference data.

\clearpage

\subsubsection{\ch{PbI2} precursor geometry}

To validate the force field in dynamical situations, we compared the finite temperature geometry of hexagonal (2H) \ch{PbI2} with experiments~\cite{schluterElectronicStructureOptical1974}. We simulated the material at \SI{300}{\K} and atmospheric pressure. First an equilibration run of \SI{50}{\ps} was done, following this equilibration run a production run of \SI{200}{\ps} was started. The lattice vectors (Table~\ref{tab:si_pbi2_geometry}) were averaged over the full duration of the production runs.

\begin{table}[htbp!]
    \caption{Comparison of lattice vectors of \ch{PbI2} from experiments~\cite{schluterElectronicStructureOptical1974} with ReaxFF simulations.}
    \label{tab:si_pbi2_geometry}
    \begin{tabular}{@{}ccc@{}}
        \toprule
                                & $a$ (\SI{}{\angstrom})  & $c$ (\SI{}{\angstrom})      \\ \midrule
        Experiments             & 4.56                    & 6.98                        \\
        ReaxFF (\SI{300}{\K})   & 4.65                    & 6.85                        \\ \bottomrule
    \end{tabular}
\end{table}

\clearpage

\subsection{Mixed halide model systems}

\begin{figure*}[htbp!]
    \includegraphics{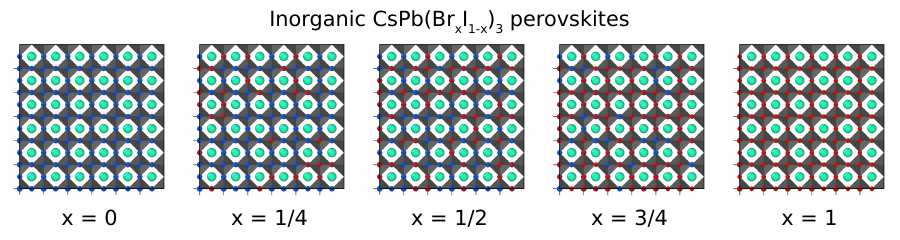}
    \caption{Mixed halide cubic \ch{CsPb(Br_{x}I_{1-x})_{3}} models created through random placement of different halide species on the inorganic sublattice. Iodide and bromide species are colored blue and red, respectively.}
    \label{fig:si_mixed_halide_systems} 
\end{figure*}

Using the $^{207}\ch{Pb}$ chemical shift to probe the local chemical environment of \ch{Pb} in inorganic halide perovskites, \citet{karmakarMechanochemicalSynthesis0D2019} established that halides are randomly distributed in the perovskite lattice. As such, we create mixed halide perovskite systems by randomly placing different halide species (i.e. \ch{I} or \ch{Br}) on the \ch{X}-site of the \ch{AMX3} perovskite lattice. By varying the number of \ch{I} and \ch{Br} species, we can control the final composition of the model system. In the creation of the structures, we ensured that an equal number of halides were substituted along each of the principal axes of the \ch{PbX6} octahedra for a homogeneous halide distribution. Examples of cubic \ch{CsPb(Br_{x}I_{1-x})_{3}} mixed halide perovskites are shown in Figure~\ref{fig:si_mixed_halide_systems}.

\clearpage

\subsection{Molecular dynamics simulations}

The ReaxFF molecular dynamics simulations were done in AMS2022~\cite{AMS2022}, using the \ch{CsPb(Br_{x}I_{1-x})_{3}} force field developed in this work. In all simulations a time step of \SI{0.25}{\fs} was used and the atom positions were written out every 400 steps (\SI{0.1}{\ps}). All simulations were carried out in an $NpT$-ensemble in which the temperature and pressure were controlled using a thermostat and barostat with damping constants of $\tau_{\mathrm{T}} = \SI{100}{\fs}$ and $\tau_{\mathrm{p}} = \SI{2500}{\fs}$, respectively. In the simulations, the initial velocities were assigned according to a Maxwell-Boltzmann distribution at the initial temperature. During the equilibration runs we made use of the Berendsen thermostat and barostat~\cite{berendsenMolecularDynamicsCoupling1984} to control the pressure and temperature. The information of the final frame (i.e. positions and velocities) was then used as the starting point of the production runs. For the production runs, we employed the Nos\'{e}-Hoover chains (NHC) thermostat ($N_{\mathrm{chain}} = 10$)~\cite{martynaNoseHooverChains1992} and Martyna-Tobias-Klein (MTK) barostat~\cite{martynaConstantPressureMolecular1994} for pressure and temperature control, respectively.

\subsubsection{Unit cell volumes}

To investigate the unit cell volumes predicted by ReaxFF simulations, \supercell{4}{4}{3} supercells of orthorhombic \ch{CsPb(Br_{x}I_{1-x})_{3}} were used. The systems were first equilibrated to the target temperature (\SI{300}{\K}) at atmospheric pressure for \SI{50}{\ps}. After these runs, production runs were started to sample the unit cell volume for \SI{200}{\ps}. The volume of the pseudocubic unit cell $V_{\mathrm{pc}}$ was obtained by dividing the average of the total system volume by the number of formula units in the supercell. The pseudocubic lattice vector then calculated, under the assumption of cubic unit cells, as $a = \sqrt[3]{V_{\mathrm{pc}}}$.

\clearpage

\subsubsection{Phase diagrams}

\begin{figure*}[htbp!]
    \includegraphics{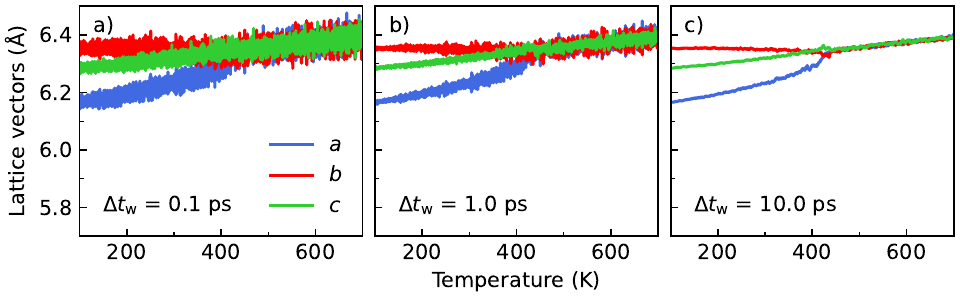}
    \caption{Effects of window averaging on the phase diagram of \ch{CsPbI3}. The investigated window widths are a) no window (\SI{0.1}{\ps}), b) \SI{1.0}{\ps} and c) \SI{10.0}{\ps}.}
    \label{fig:si_pd_window_averaging}
\end{figure*}

\begin{figure*}[htbp!]
    \includegraphics{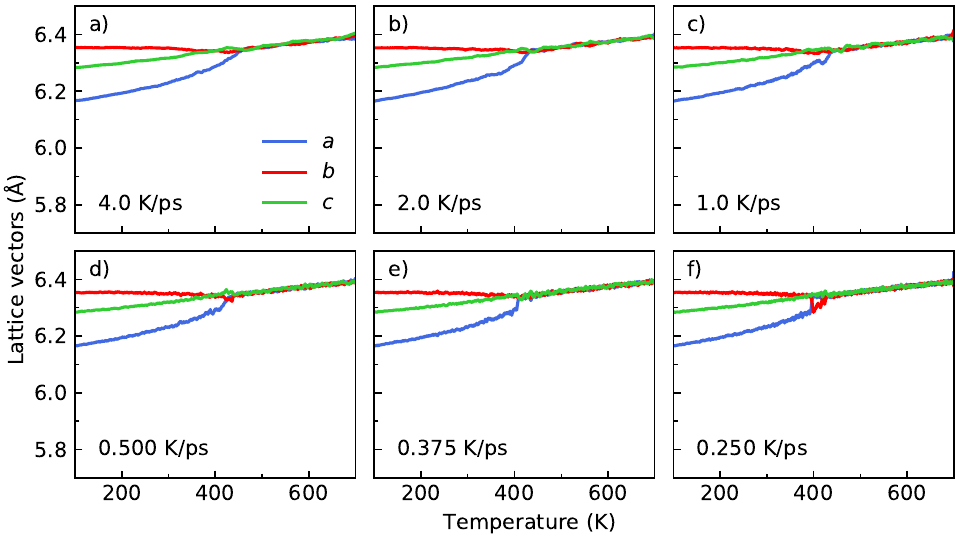}
    \caption{Effects of heating rate on the phase diagram of \ch{CsPbI3}. The investigated heating rates are a) \SI{4.0}{K/ps}, b) \SI{2.0}{K/ps}, c) \SI{1.0}{K/ps}, d) \SI{0.500}{K/ps} e) \SI{0.375}{K/ps} and f) \SI{0.250}{K/ps}.}
    \label{fig:si_pd_heating_rate}
\end{figure*}

\begin{figure*}[htbp!]
    \includegraphics{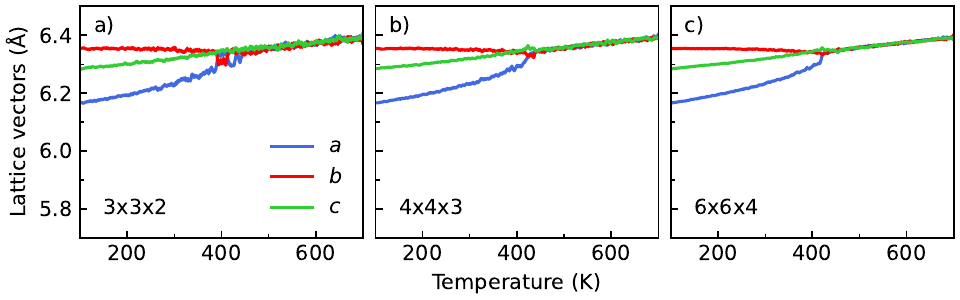}
    \caption{Effects of supercell size on the phase diagram of \ch{CsPbI3}. The investigated supercell sizes are a) \supercell{3}{3}{2}, b) \supercell{4}{4}{3} and c) \supercell{6}{6}{4}.}
    \label{fig:si_pd_cell_size}
\end{figure*}

\begin{figure*}[htbp!]
    \includegraphics{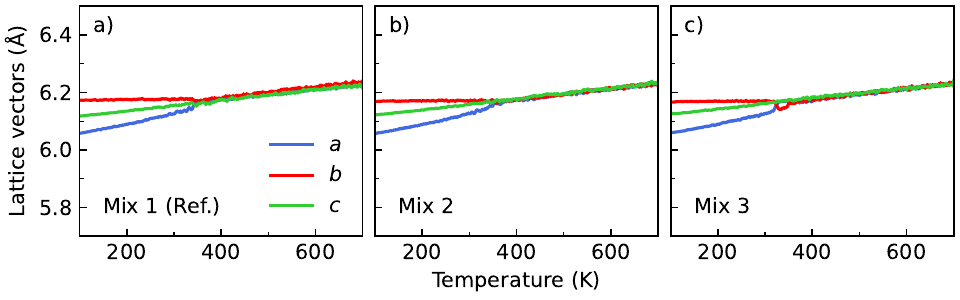}
    \caption{Effects of differences in halide mixing on the phase diagram of \ch{CsPb(Br_{1/4}I_{3/4})_{3}}. Three model systems with identical halide ratios are compared in the subfigures.}
    \label{fig:si_pd_halide_mixing}
\end{figure*}

Phase diagrams were created with \supercell{4}{4}{3} supercells of orthorhombic \ch{CsPb(Br_{x}I_{1-x})_{3}}. All compositions were initially equilibrated to \SI{100}{\K} at atmospheric pressure for \SI{25}{\ps}. All systems were then gradually heated from \SI{100}{\K} to \SI{700}{\K} during a \SI{1.2}{\ns} simulation, resulting in a heating rate of \SI{0.5}{\K\per\ps}. Pseudocubic lattice vectors $a$, $b$ and $c$ were consequently obtained by converting the orthorhombic cell vectors $a_{\mathrm{o}}$, $b_{\mathrm{o}}$ and $c_{\mathrm{o}}$ as: $a = a_{\mathrm{o}} / \sqrt{2}$, $b = b_{\mathrm{o}} / \sqrt{2}$ and $c = c_{\mathrm{o}} / 2$. A window averaging with a window width of $\Delta t_{\mathrm{w}} = \SI{10.0}{\ps}$ was used to smooth the fluctuations in phase diagrams. Figure~\ref{fig:si_pd_window_averaging} demonstrates that the smoothing of this data does not impact the location of the phase transitions in \ch{CsPbI3}. Moreover, sensitivity tests of the phase diagrams for the heating rate (Figure~\ref{fig:si_pd_heating_rate}), supercell size (Figure~\ref{fig:si_pd_cell_size}) and halide mixing (Figure~\ref{fig:si_pd_halide_mixing}) show the robust behavior of the phase diagrams of the inorganic perovskites. We note that hysteresis effects have been observed during the rapid cooling of \ch{CsPbI3} before~\cite{polsAtomisticInsightsDegradation2021}, yet, by focusing on the phase diagrams from the heating of the inorganic perovskites, such hysteresis effects do not have to be taken into account.

\clearpage

\subsubsection{Homogeneously mixed systems}

\begin{figure*}[htbp!]
    \includegraphics{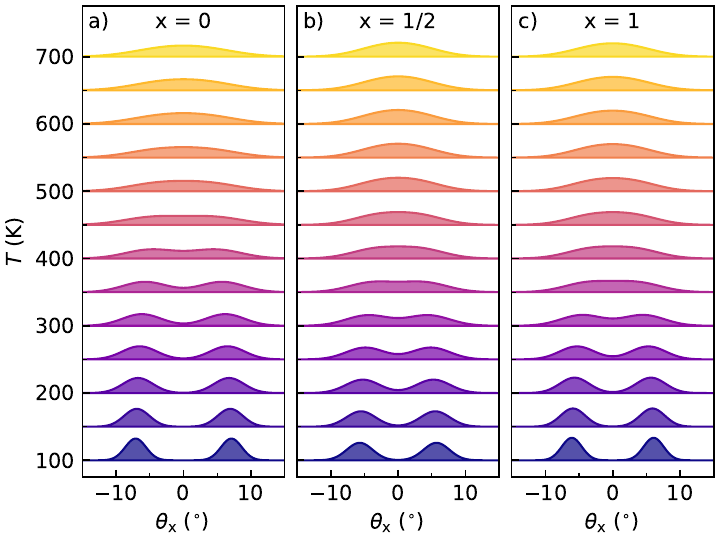}
    \caption{Temperature evolution of the octahedral orientation $\thetax$ for \ch{CsPb(Br_{x}I_{1-x})_{3}} perovskites with compositions a) x = 0, b) x = 1/2 and c) x = 1.}
    \label{fig:si_octahedral_distributions_x}
\end{figure*}

\begin{figure*}[htbp!]
    \includegraphics{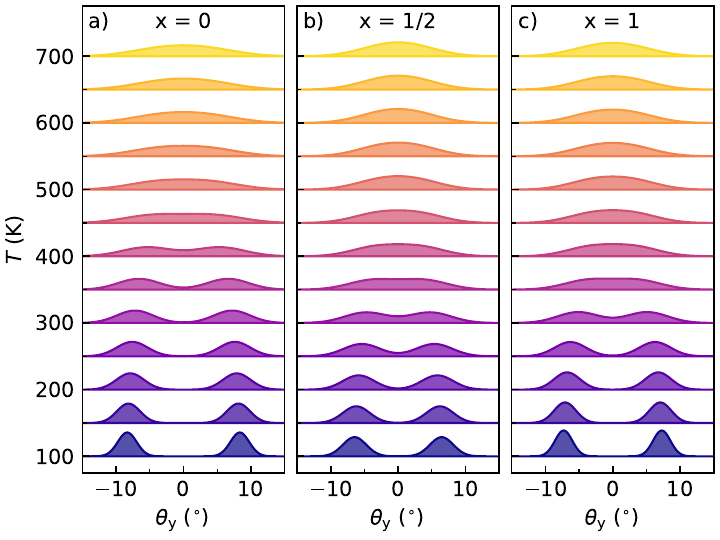}
    \caption{Temperature evolution of the octahedral orientation $\thetay$ for \ch{CsPb(Br_{x}I_{1-x})_{3}} perovskites with compositions a) x = 0, b) x = 1/2 and c) x = 1.}
    \label{fig:si_octahedral_distributions_y}
\end{figure*}

The octahedral dynamics in homogeneously mixed systems were analyzed in \supercell{4}{4}{3} orthorhombic supercells. The model systems were equilibrated to \SI{75}{\K} at atmospheric pressure for \SI{50}{\ps}. During the production runs (\SI{1.3}{\ns}) the temperature was gradually increased to \SI{725}{\K}, resulting in a heating rate of \SI{0.5}{\K\per\ps}. To obtain octahedral distributions, we divided the trajectory in sections of \SI{100}{\ps}, each representing a temperature window of \SI{50}{\K}. The distributions were analyzed using the functional forms described in SI Note~\ref{sec:octahedral_orientation}. The temperature evolution of the average octahedral tilt angles was obtained by dividing the trajectory into sections of \SI{20}{\ps} (i.e. \SI{10}{\K} windows). The temperature evolution of $\thetax$ and $\thetay$ can be found in Figure~\ref{fig:si_octahedral_distributions_x} and Figure~\ref{fig:si_octahedral_distributions_y}, respectively.

\clearpage

\subsubsection{Dilute mixed systems}

To investigate the local effects of halide substitutions in the inorganic perovskites, we used \supercell{4}{4}{3} orthorhombic supercells. From these cells we randomly selected a \ch{Pb} species and substituted one or two of the neighboring halides species on either the axial or equatorial sites. The model systems were then equilibrated to the target temperature of \SI{300}{\K} at atmospheric pressure for \SI{50}{\ps}, followed by a production run of \SI{500}{\ps} to sample the octahedral dynamics. The average tilt angles $\langle {\thetai} \rangle$ and standard deviations in those tilt angles $\sigmai$ (i = x, y, z) as obtained during the full production runs are shown in Table~\ref{tab:si_tilting_distributions_dilute}.

\begin{table}[htbp!]
    \caption{Averages $\langle {\thetai} \rangle$ and standard deviations $\sigma_{\mathrm{i}}$ (i = x, y, z) of the octahedral tilting of \ch{PbX6} octahedra in pure \ch{CsPbI3}, pure \ch{CsPbBr3} and substituted \ch{CsPbI3} at \SI{300}{\K}.}
    \label{tab:si_tilting_distributions_dilute}
    \begin{tabular}{@{}cccccccc@{}}
        \toprule
        Octahedral type         & $\langle {\thetax} \rangle$ $\left( ^{\circ} \right)$    & $\langle {\thetay} \rangle$ $\left( ^{\circ} \right)$    & $\langle {\thetaz} \rangle$ $\left( ^{\circ} \right)$    & $\sigma_{\mathrm{x}}$ $\left( ^{\circ} \right)$     & $\sigma_{\mathrm{y}}$ $\left( ^{\circ} \right)$     & $\sigma_{\mathrm{z}}$ $\left( ^{\circ} \right)$     \\ \midrule
        \ch{CsPbI3}             & 6.20                  & 7.22                  & 8.93                  & 2.88                      & 2.66                      & 2.51                      \\
        Axial substitution      & 5.67                  & 5.55                  & 8.50                  & 2.34                      & 2.21                      & 2.61                      \\
        Equatorial substitution & 4.80                  & 5.52                  & 7.11                  & 2.76                      & 2.48                      & 2.29                      \\ 
        \ch{CsPbBr3}            & 4.59                  & 4.57                  & 4.48                  & 3.41                      & 3.39                      & 3.38                      \\ \bottomrule
    \end{tabular}
\end{table}

\clearpage

\subsection{Octahedral orientation}
\label{sec:octahedral_orientation}

\subsubsection{Determination of orientation}

\begin{figure*}[htbp!]
    \includegraphics{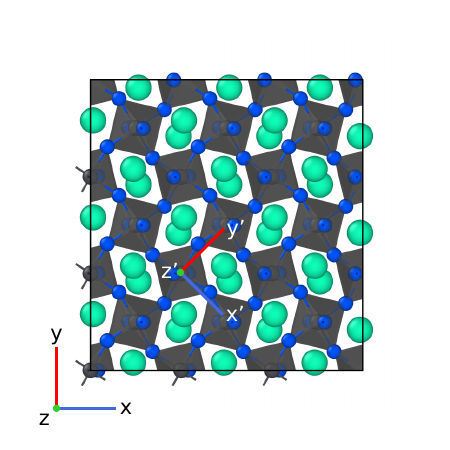}
    \caption{Rotated coordinate axis ($x'y'z'$) used to determine the octahedral orientation for \ch{PbI6} octahedra in orthorhombic \ch{CsPbI3}. The original coordinate axis ($xyz$) is also shown.}
    \label{fig:si_axis_orientation}
\end{figure*}

We determine the orientation of the octahedra in the inorganic \ch{PbX6} framework of the inorganic halide perovskites using polyhedral template matching (PTM) as implemented in \texttt{OVITO}~\cite{larsenRobustStructuralIdentification2016, stukowskiVisualizationAnalysisAtomistic2009}. The neighborhood of \ch{PbX6} octahedra closely resembles the \texttt{Simple Cubic} environment. Using a root-mean-square deviation (RMSD) cutoff for PTM of $0.2$, a robust identification of the octahedra is realized in both pure and mixed halide perovskites. The orientation of the octahedra is consequently given in terms of the Euler angles ($\thetax$, $\thetay$ and $\thetaz$) using the $x$-$y$-$z$ order Tait-Bryan convention with extrinsic rotations (see Figure 4a). Whenever the orthorhombic structure of \ch{CsPbX3} perovskites is used, we rotate the coordinate axis with respect to which the octahedral orientation is determined with \ang{45} around the $z$-axis. The rotated coordinate axis is shown in Figure~\ref{fig:si_axis_orientation}.

\clearpage

\subsubsection{Analysis of orientation}

The location and spread of the octahedral tilting distributions can consequently be analyzed using Gaussian distributions. Due to an out-of-phase tilting pattern of the octahedra in low-temperature \ch{CsPbX3} phases, the octahedral angles have a bimodal tilting distribution. Such bimodal distributions can be analyzed using a symmetric Gaussian function around zero~\cite{wiktorQuantifyingDynamicTilting2023} of the form
\begin{equation}
    \label{eq:double_gaussian}
    g_{2}(\thetai) = \frac{1}{2 \sigmai \sqrt{2 \pi}} \left( \exp \left[ -\frac{\left( \thetai - \langle \thetai \rangle \right)^{2}}{2 \sigmai^{2}} \right] + \exp \left[ -\frac{\left( \thetai + \langle \thetai \rangle \right)^{2}}{2 \sigmai^{2}} \right] \right)
\end{equation}
where $\thetai$ is an arbitrary octahedral angle (i = x, y, z),  $\langle \thetai \rangle$ is the average tilting angle and $\sigmai$ is the standard deviation in the tilting angle. Whenever the tilting distribution of a single \ch{PbX6} octahedron is investigated, a single Gaussian function is used
\begin{equation}
    \label{eq:single_gaussian}
    g_{1}(\thetai) = \frac{1}{\sigmai \sqrt{2 \pi}} \exp \left[ -\frac{\left( \thetai - \langle \thetai \rangle \right)^{2}}{2 \sigmai^{2}} \right].
\end{equation}

\begin{figure*}[htbp!]
    \includegraphics{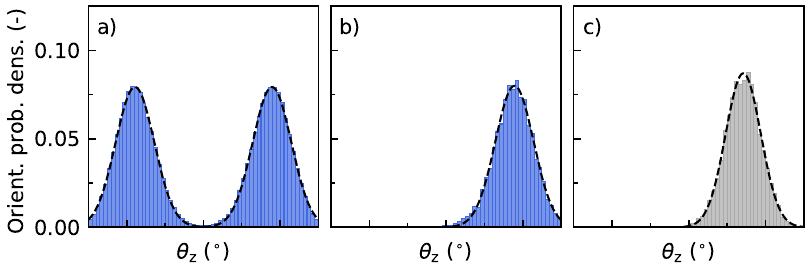}
    \caption{Comparison of octahedral tilting distributions of $\thetaz$ with fits with Gaussian distributions. Tilting distributions of a) \ch{PbI6} octahedra in pure \ch{CsPbI3}, b) a single \ch{PbI6} octahedron in \ch{CsPbI3} and c) a single octahedron in \ch{CsPbI3} that is doubly substituted by \ch{Br} in the equatorial direction. In all plots, the dashed black line shows the Gaussian fits, whereas the simulated distributions are depicted with colored bars.}
    \label{fig:si_octahedral_distribution_analysis}
\end{figure*}

To validate the use of the aforementioned Gaussian distributions for analysis of the octahedral distributions, we compare fits of $g_{1}$ and $g_{2}$ to octahedral distributions from simulations. Figure~\ref{fig:si_octahedral_distribution_analysis} shows that the Gaussian distributions match well with the distributions obtained from ReaxFF simulations at \SI{300}{\K}. Specifically, we find that the double Gaussian distribution describes the collective orientation of \ch{PbI6} octahedra in pure \ch{CsPbI3} (Figure~\ref{fig:si_octahedral_distribution_analysis}a). Moreover, the tilting distributions of both individual pure and substituted octahedra in inorganic perovskites (Figure~\ref{fig:si_octahedral_distribution_analysis}b and Figure~\ref{fig:si_octahedral_distribution_analysis}c) are well-described by a single Gaussian distribution.

\clearpage

\subsection{Substitution propagation}

\subsubsection{Single substitution}

\begin{figure*}[htbp!]
    \includegraphics{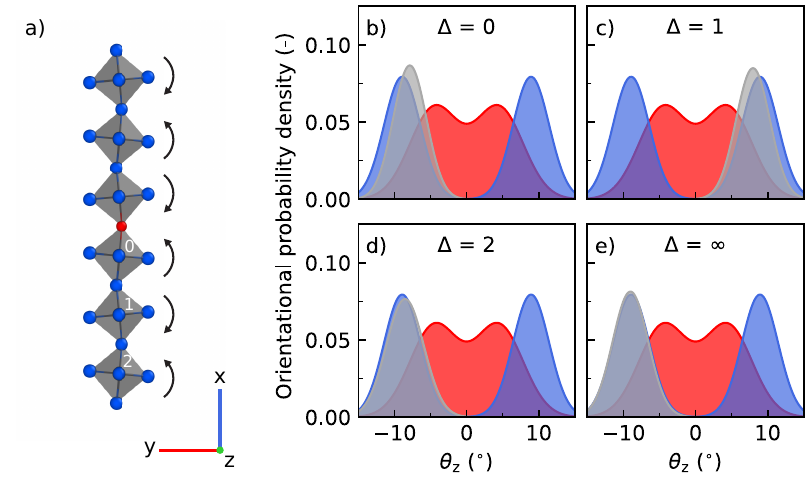}
    \caption{a) Chain of \ch{PbX6} octahedra with a single substitution. The numbers in the octahedra indicate the distance relative to the substituted octahedron. Distribution of $\thetaz$ of b) substituted octahedron ($\Delta = 0$), c) direct neighbor of the substituted octahedron ($\Delta = 1$), d) octahedron two sites away from the substituted octahedron ($\Delta = 2$) and e) reference octahedron very far away from the halide substitution ($\Delta = \infty$). The tilting distributions of the investigated octahedra are shown in gray, and those for \ch{CsPbI3} and \ch{CsPbBr3} in blue and red, respectively.}
    \label{fig:si_single_substitution}
\end{figure*}

To highlight that the substitution of a single halide is enough to impact the dynamics of neighboring octahedra, we assess the octahedral dynamics in \ch{CsPbI3} with a single \ch{Br} substitution in the equatorial direction in Figure~\ref{fig:si_single_substitution}. We find a similar strain effect for such single halide substitutions as for double halide substitutions. Interestingly, the effect appears to be slightly smaller for single substitutions than for double substitutions as shown in Table~\ref{tab:si_neighbor_distributions}, hinting at an additive nature of this strain effect.

\begin{table}[htbp!]
    \caption{Averages $\langle {\thetaz} \rangle$ and standard deviations $\sigma_{\mathrm{z}}$ in the octahedral tilting of \ch{PbX6} octahedra in equatorially substituted \ch{CsPbI3} at \SI{300}{\K} in the dilute limit. The distance from the substituted octahedron is expressed in $\Delta$.}
    \label{tab:si_neighbor_distributions}
    \begin{tabular}{@{}ccccc@{}}
        \toprule
        \multirow{2}{*}{$\Delta$}   & \multicolumn{2}{c}{2 substitutions}                                                                            & \multicolumn{2}{c}{1 substitution}                                                                       \\ 
                                    & $\langle {\thetaz} \rangle$ $\left( ^{\circ} \right)$     & $\sigma_{\mathrm{z}}$ $\left( ^{\circ} \right)$   & $\langle {\thetaz} \rangle$ $\left( ^{\circ} \right)$     & $\sigma_{\mathrm{z}}$ $\left( ^{\circ} \right)$   \\ \midrule
        0                           & 7.11                                                      & 2.29                                              & 7.88                                                      & 2.30                                              \\
        1                           & 7.50                                                      & 2.46                                              & 7.96                                                      & 2.35                                              \\
        2                           & 7.70                                                      & 2.44                                              & 8.54                                                      & 2.61                                              \\
        $\infty$                    & 9.01                                                      & 2.49                                              & 9.02                                                      & 2.45                                              \\ \bottomrule
    \end{tabular}
\end{table}

\subsubsection{Propagation distance}

\begin{figure*}[htbp!]
    \includegraphics{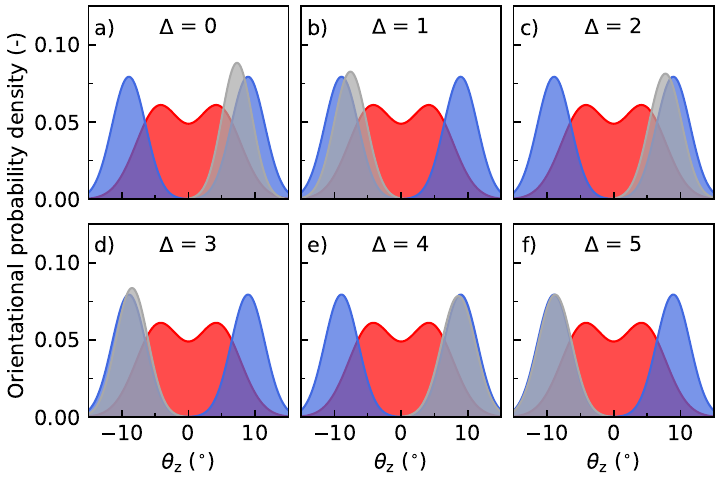}
    \caption{Octahedral distributions of $\thetaz$ for the a) substituted octahedron ($\Delta = 0$) and octahedra b) one ($\Delta = 1$), c) two ($\Delta = 2$), d) three ($\Delta = 3$), e) four ($\Delta = 4$) and f) five ($\Delta = 5$) octahedra away from the substituted octahedron as indicated by $\Delta$.}
    \label{fig:si_propagation_distance}
\end{figure*}

To determine the range of the strain effect of halide substitutions, we assess the octahedral orientations around a double equatorial halide substitution in a larger orthorhombic supercell (\supercell{6}{6}{4}). The resulting octahedral distributions are shown in Figure~\ref{fig:si_propagation_distance}, where we used the same naming convention as in Figure 6, with $\Delta$ indicating the distance of the octahedra away from the substitution. The overview of the tilt angles (Table~\ref{tab:si_propagation_distance}) demonstrates that the propagation of the strain effect diminishes significantly for octahedra spaced further than three sites away from the substitution ($\Delta > 3$), indicating a propagation distance of \SI{2}{\nm} for this strain effect in the direction of the substitutions.

\begin{table}[htbp!]
    \caption{Averages $\langle {\thetaz} \rangle$ and standard deviations $\sigma_{\mathrm{z}}$ in the octahedral tilting of \ch{PbX6} octahedra in \supercell{6}{6}{4} supercells of \ch{CsPbI3} at \SI{300}{\K}. The distance from the equatorially substituted octahedron is expressed in $\Delta$.}
    \label{tab:si_propagation_distance}
    \begin{tabular}{@{}ccc@{}}
        \toprule
        $\Delta$                    & $\langle {\thetaz} \rangle$ $\left( ^{\circ} \right)$     & $\sigma_{\mathrm{z}}$ $\left( ^{\circ} \right)$   \\ \midrule
        0                           & 7.28                                                      & 2.26                                              \\
        1                           & 7.60                                                      & 2.41                                              \\
        2                           & 7.75                                                      & 2.45                                              \\
        3                           & 8.46                                                      & 2.39                                              \\
        4                           & 8.59                                                      & 2.54                                              \\
        5                           & 8.78                                                      & 2.52                                              \\
        $\infty$                    & 8.82                                                      & 2.53                                              \\ \bottomrule
    \end{tabular}
\end{table}

\subsubsection{Perpendicular propagation}

\begin{figure*}[htbp!]
    \includegraphics{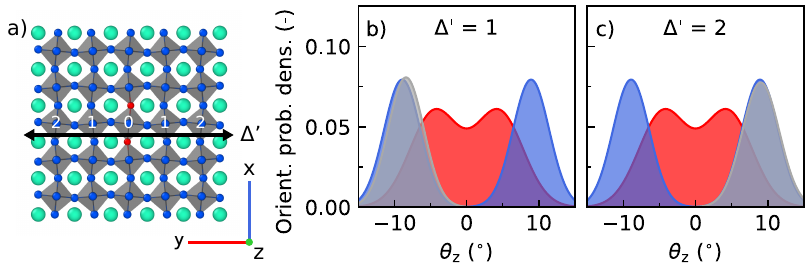}
    \caption{a) Schematic overview of the halide substitution and the investigated octahedra. The numbers in the octahedra indicate the distance relative to the substituted octahedron. Distribution of $\thetaz$ of b) direct neighbor of the substituted octahedron ($\Delta' = 1$) and c) octahedron two sites away from the substituted octahedron ($\Delta' = 2$). The tilting distributions of the investigated octahedra are shown in gray, and those for \ch{CsPbI3} and \ch{CsPbBr3} in blue and red, respectively.}
    \label{fig:si_perpendicular_propagation}
\end{figure*}

To illustrate that halide substitutions do not solely affect octahedra in the direction of the substitution, we show the changes in the octahedral tilting for octahedra arranged perpendicular from the substitution direction in Figure~\ref{fig:si_perpendicular_propagation}. Although the effect is significantly smaller for such octahedra, it results in shifts of about \SI{0.6}{\degree} for the direct neighbors as shown in Table~\ref{tab:si_perpendicular_propagation}. As is demonstrated by the negligible impact on octahedra spaced two sites away from the substitution, this effect has a considerably shorter range in direction perpendicular to the substitution ($< \SI{1.0}{\nm}$) than in the direction of the substitution (\SI{2}{\nm}).

\begin{table}[htbp!]
    \caption{Averages $\langle {\thetaz} \rangle$ and standard deviations $\sigma_{\mathrm{z}}$ in the octahedral tilting of \ch{PbX6} octahedra in substituted \ch{CsPbI3} at \SI{300}{\K} in the dilute limit. The perpendicular distance from the substituted octahedron is expressed in $\Delta'$. The tilting distributions of the investigated octahedra are shown in gray, and those for \ch{CsPbI3} and \ch{CsPbBr3} in blue and red, respectively.}
    \label{tab:si_perpendicular_propagation} 
    \begin{tabular}{@{}ccc@{}}
        \toprule
        $\Delta'$       & $\langle {\thetaz} \rangle$ $\left( ^{\circ} \right)$     & $\sigma_{\mathrm{z}}$ $\left( ^{\circ} \right)$   \\ \midrule
        0               & 7.11                                                      & 2.29                                              \\
        1               & 8.40                                                      & 2.47                                              \\
        2               & 9.05                                                      & 2.57                                              \\
        $\infty$        & 9.01                                                      & 2.49                                              \\ \bottomrule
    \end{tabular}
\end{table}

\clearpage

\subsection{Tolerance factors}

The Goldschmidt tolerance factor is an empirical factor to predict the distortion of an \ch{AMX3} perovskite crystal structure~\cite{goldschmidtGesetzeKrystallochemie1926} as
\begin{equation}
    \label{eq:tolerance_factors}
    t = \frac{r_{\mathrm{A}} + r_{\mathrm{X}}}{\sqrt{2} \left( r_{\mathrm{M}} + r_{\mathrm{X}} \right)}
\end{equation}
where $r_{\mathrm{A}}$, $r_{\mathrm{M}}$ and $r_{\mathrm{X}}$ are the ionic radii of \ch{A}, \ch{M} and \ch{X} ions, respectively. Materials with a tolerance factor of $0.90 - 1.00$ have a cubic perovskite structure, lower tolerance factors ($0.71 - 0.90$) indicate an orthorhombic perovskite structure with tilted octahedra and materials with very high ($> 1.00$) or very low ($<0.71$) tolerance factors suggest the formation of nonperovskite structures.

Using the ionic radii from \citet{shannonRevisedEffectiveIonic1976} shown in Table~\ref{tab:si_ionic_radii}, the Goldschmidt tolerance factors of \ch{CsPbX3} perovskites can be calculated using equation~\ref{eq:tolerance_factors}. \ch{CsPbBr3} (0.815) has a slightly higher tolerance factor than \ch{CsPbI3} (0.807), indicating it has a slightly less distorted perovskite structure.

\begin{table}[htbp!]
    \caption{Ionic radii reported by \citet{shannonRevisedEffectiveIonic1976}.}
    \label{tab:si_ionic_radii} 
    \begin{tabular}{@{}cc@{}}
        \toprule
        Element         & $r_{\mathrm{ion}}$ (\SI{}{\angstrom})     \\ \midrule
        \ch{Cs}         & 1.67                                      \\
        \ch{Pb}         & 1.19                                      \\
        \ch{Br}         & 1.96                                      \\
        \ch{I}          & 2.20                                      \\ \bottomrule
    \end{tabular}
\end{table}


\clearpage

\bibliography{supporting}